\input amssym.def
\input epsf

% Page layout

\magnification=\magstephalf
\hsize=14.0 true cm
\vsize=19 true cm
\hoffset=1.0 true cm
\voffset=2.0 true cm

\abovedisplayskip=12pt plus 3pt minus 3pt
\belowdisplayskip=12pt plus 3pt minus 3pt
\parindent=1.0em

% Fonts

\font\sixrm=cmr6
\font\eightrm=cmr8
\font\ninerm=cmr9

\font\sixi=cmmi6
\font\eighti=cmmi8
\font\ninei=cmmi9

\font\sixsy=cmsy6
\font\eightsy=cmsy8
\font\ninesy=cmsy9

\font\sixbf=cmbx6
\font\eightbf=cmbx8
\font\ninebf=cmbx9

\font\eightit=cmti8
\font\nineit=cmti9

\font\eightsl=cmsl8
\font\ninesl=cmsl9

\font\sixss=cmss8 at 8 true pt
\font\sevenss=cmss9 at 9 true pt
\font\eightss=cmss8
\font\niness=cmss9
\font\tenss=cmss10

 at 12 true pt
 at 12 true pt
\font\bigbf=cmbx10 at 12 true pt

 at 14 true pt
 at 16 true pt
 at 14 true pt

\catcode`@=11
\newfam\ssfam

\def\tenpoint{\def\rm{\fam0\tenrm}%
    \textfont0=\tenrm \scriptfont0=\sevenrm \scriptscriptfont0=\fiverm
    \textfont1=\teni  \scriptfont1=\seveni  \scriptscriptfont1=\fivei
    \textfont2=\tensy \scriptfont2=\sevensy \scriptscriptfont2=\fivesy
    \textfont3=\tenex \scriptfont3=\tenex   \scriptscriptfont3=\tenex
    \textfont\itfam=\tenit                  \def\it{\fam\itfam\tenit}%
    \textfont\slfam=\tensl                  \def\sl{\fam\slfam\tensl}%
    \textfont\bffam=\tenbf \scriptfont\bffam=\sevenbf
    \scriptscriptfont\bffam=\fivebf
                                            \def\bf{\fam\bffam\tenbf}%
    \textfont\ssfam=\tenss \scriptfont\ssfam=\sevenss
    \scriptscriptfont\ssfam=\sevenss
                                            \def\ss{\fam\ssfam\tenss}%
    \normalbaselineskip=13pt
    \setbox\strutbox=\hbox{\vrule height8.5pt depth3.5pt width0pt}%
    \let\big=\tenbig
    \normalbaselines\rm}

\def\ninepoint{\def\rm{\fam0\ninerm}%
    \textfont0=\ninerm      \scriptfont0=\sixrm
                            \scriptscriptfont0=\fiverm
    \textfont1=\ninei       \scriptfont1=\sixi
                            \scriptscriptfont1=\fivei
    \textfont2=\ninesy      \scriptfont2=\sixsy
                            \scriptscriptfont2=\fivesy
    \textfont3=\tenex       \scriptfont3=\tenex
                            \scriptscriptfont3=\tenex
    \textfont\itfam=\nineit \def\it{\fam\itfam\nineit}%
    \textfont\slfam=\ninesl \def\sl{\fam\slfam\ninesl}%
    \textfont\bffam=\ninebf \scriptfont\bffam=\sixbf
                            \scriptscriptfont\bffam=\fivebf
                            \def\bf{\fam\bffam\ninebf}%
    \textfont\ssfam=\niness \scriptfont\ssfam=\sixss
                            \scriptscriptfont\ssfam=\sixss
                            \def\ss{\fam\ssfam\niness}%
    \normalbaselineskip=12pt
    \setbox\strutbox=\hbox{\vrule height8.0pt depth3.0pt width0pt}%
    \let\big=\ninebig
    \normalbaselines\rm}

\def\eightpoint{\def\rm{\fam0\eightrm}%
    \textfont0=\eightrm      \scriptfont0=\sixrm
                             \scriptscriptfont0=\fiverm
    \textfont1=\eighti       \scriptfont1=\sixi
                             \scriptscriptfont1=\fivei
    \textfont2=\eightsy      \scriptfont2=\sixsy
                             \scriptscriptfont2=\fivesy
    \textfont3=\tenex        \scriptfont3=\tenex
                             \scriptscriptfont3=\tenex
    \textfont\itfam=\eightit \def\it{\fam\itfam\eightit}%
    \textfont\slfam=\eightsl \def\sl{\fam\slfam\eightsl}%
    \textfont\bffam=\eightbf \scriptfont\bffam=\sixbf
                             \scriptscriptfont\bffam=\fivebf
                             \def\bf{\fam\bffam\eightbf}%
    \textfont\ssfam=\eightss \scriptfont\ssfam=\sixss
                             \scriptscriptfont\ssfam=\sixss
                             \def\ss{\fam\ssfam\eightss}%
    \normalbaselineskip=10pt
    \setbox\strutbox=\hbox{\vrule height7.0pt depth2.0pt width0pt}%
    \let\big=\eightbig
    \normalbaselines\rm}

\def\tenbig#1{{\hbox{$\left#1\vbox to8.5pt{}\right.\n@space$}}}
\def\ninebig#1{{\hbox{$\textfont0=\tenrm\textfont2=\tensy
                       \left#1\vbox to7.25pt{}\right.\n@space$}}}
\def\eightbig#1{{\hbox{$\textfont0=\ninerm\textfont2=\ninesy
                       \left#1\vbox to6.5pt{}\right.\n@space$}}}

\font\sectionfont=cmbx10
\font\subsectionfont=cmti10

\def\figurecaptionfont{\ninepoint}
\def\tablecaptionfont{\ninepoint}
\def\footnotefont{\eightpoint}

% New count registers

\newcount\equationno
\newcount\bibitemno
\newcount\figureno
\newcount\tableno

\equationno=0
\bibitemno=0
\figureno=0
\tableno=0
%\advance\pageno by -1

% Footline

\footline={\ifnum\pageno=0{\hfil}\else
{\hss\rm\the\pageno\hss}\fi}

% Section macro

\def\section #1. #2 \par
{\vskip0pt plus .20\vsize\penalty-100 \vskip0pt plus-.20\vsize
\vskip 1.6 true cm plus 0.2 true cm minus 0.2 true cm
\global\def\equationlabel{#1}
\global\equationno=0
\leftline{\sectionfont #1. #2}\par
\immediate\write\terminal{Section #1. #2}
\vskip 0.7 true cm plus 0.1 true cm minus 0.1 true cm
\noindent}

% Subsection macro

\def\subsection #1 \par
{\vskip0pt plus 0.8 true cm\penalty-50 \vskip0pt plus-0.8 true cm
\vskip2.5ex plus 0.1ex minus 0.1ex
\leftline{\subsectionfont #1}\par
\immediate\write\terminal{Subsection #1}
\vskip1.0ex plus 0.1ex minus 0.1ex
\noindent}

% Appendix macro

\def\appendix #1. #2 \par
{\vskip0pt plus 0.8 true cm\penalty-50 \vskip0pt plus-0.8 true cm
\vskip2.5ex plus 0.1ex minus 0.1ex
\global\def\equationlabel{\hbox{\rm#1}}
\global\equationno=0
\leftline{\subsectionfont Appendix #1. #2}\par
\immediate\write\terminal{Appendix #1. #2}
\vskip1.0ex plus 0.1ex minus 0.1ex
\noindent}

% Displayed equations

\def\equation#1{$$\displaylines{\qquad #1}$$}
\def\enum{\global\advance\equationno by 1
\hfill\llap{(\equationlabel.\the\equationno)}}
\def\noenum{\hfill}
\def\next#1{\cr\noalign{\vskip#1}\qquad}

% Bibliography macro, references

\def\ifundefined#1{\expandafter\ifx\csname#1\endcsname\relax}

\def\ref#1{\ifundefined{#1}?\immediate\write\terminal{unknown reference
on page \the\pageno}\else\csname#1\endcsname\fi}

\newwrite\terminal
\newwrite\bibitemlist

\def\bibitem#1#2\par{\global\advance\bibitemno by 1
\immediate\write\bibitemlist{\string\def
\expandafter\string\csname#1\endcsname
{\the\bibitemno}}
\item{[\the\bibitemno]}#2\par}

\def\beginbibliography{
\vskip0pt plus .15\vsize\penalty-100 \vskip0pt plus-.15\vsize
\vskip 1.2 true cm plus 0.2 true cm minus 0.2 true cm
\leftline{\sectionfont References}\par
\immediate\write\terminal{References}
\immediate\openout\bibitemlist=biblist
\frenchspacing\parindent=1.8em
\vskip 0.5 true cm plus 0.1 true cm minus 0.1 true cm}

\def\endbibliography{
\immediate\closeout\bibitemlist
\nonfrenchspacing\parindent=1.0em}

% Figure and table captions

\def\figurecaption#1{\global\advance\figureno by 1
\narrower\figurecaptionfont
Fig.~\the\figureno. #1}

\def\tablecaption#1{\global\advance\tableno by 1
\vbox to 0.5 true cm { }
\centerline{\tablecaptionfont%
Table~\the\tableno. #1}
\vskip-0.4 true cm}

\tenpoint

\immediate\openin\bibitemlist=biblist
\ifeof\bibitemlist\immediate\closein\bibitemlist
\else\immediate\closein\bibitemlist
\input biblist \fi

% current year and month

\def\thismonth{\ifcase\month\or
January\or February\or March\or April\or May\or June\or
July\or August\or September\or October\or November\or December\fi}

% Definitions and abbreviations

% Roman letters in math formulae

\def\rmd{{\rm d}}
\def\rmD{{\rm D}}
\def\rme{{\rm e}}
\def\rmO{{\rm O}}

% Real and integer numbers

\def\gz{{\Bbb Z}}
\def\Im{{\rm Im}\,}

% Special relations and symbols

\def\proof{\noindent{\sl Proof:}\kern0.6em}

\def\frac#1#2{\hbox{$#1\over#2$}}
\def\dual{\mathstrut^*\kern-0.1em}

\def\ring{\mathaccent"7017}
\def\lvec#1{\setbox0=\hbox{$#1$}
    \setbox1=\hbox{$\scriptstyle\leftarrow$}
    #1\kern-\wd0\smash{
    \raise\ht0\hbox{$\raise1pt\hbox{$\scriptstyle\leftarrow$}$}}
    \kern-\wd1\kern\wd0}
\def\rvec#1{\setbox0=\hbox{$#1$}
    \setbox1=\hbox{$\scriptstyle\rightarrow$}
    #1\kern-\wd0\smash{
    \raise\ht0\hbox{$\raise1pt\hbox{$\scriptstyle\rightarrow$}$}}
    \kern-\wd1\kern\wd0}

% Lattice derivatives

\def\nab#1{{\nabla_{#1}}}
\def\nabstar#1{{\nabla\kern0.5pt\smash{\raise 4.5pt\hbox{$\ast$}}
               \kern-5.5pt_{#1}}}
\def\drv#1{{\partial_{#1}}}
\def\drvstar#1{{\partial\kern0.5pt\smash{\raise 4.5pt\hbox{$\ast$}}
               \kern-6.0pt_{#1}}}

\def\ldrvstar#1{{\lvec{\,\partial}\kern-0.5pt\smash{\raise 4.5pt\hbox{$\ast$}}
               \kern-5.0pt_{#1}}}

% Units

% Constants

% Fields

\def\psibar{\overline{\psi}}

% Dirac matrices

\def\dirac#1{\gamma_{#1}}
\def\diracstar#1#2{
    \setbox0=\hbox{$\gamma$}\setbox1=\hbox{$\gamma_{#1}$}
    \gamma_{#1}\kern-\wd1\kern\wd0
    \smash{\raise4.5pt\hbox{$\scriptstyle#2$}}}
\def\dirachat{\hat{\gamma}_5}

% Gauge group

\def\tr{{\rm tr}}
\def\Tr{{\rm Tr}}
\def\Ad{{\rm Ad}\kern0.1em}
\def\d#1{d^{#1}_{\hbox{$\scriptstyle\kern-0.5pt R\scriptfont1=\sixi$}}}

% Lattice, action, field variations

\def\Seff{S_{\rm eff}}
\def\SF{S_{\rm F}}
\def\SG{S_{\rm G}}
\def\Phat{\hat{P}}
\def\L{{\frak L}}
\def\anomaly{{\cal A}}
\def\Nf{N_{\rm f}}
\rightline{CERN-TH/2001-031}

\vskip 1.4 true cm 
\centerline
{\bigbf  Chiral gauge theories revisited}

\vskip 0.8 true cm
\centerline{\it Lectures given at the International School of
Subnuclear Physics} 
\vskip1ex
\centerline{\it Erice, 27 August -- 5 September 2000}

\vskip 0.8 true cm
\centerline{\bf Martin L\"uscher\kern1pt%
\footnote{${\vrule height7.0pt depth1.5pt width0pt}^{\ast}$}{\footnotefont% 
On leave from Deutsches Elektronen-Synchrotron DESY, 
D-22603 Hamburg, Germany}
}
\vskip2ex
\centerline{\it CERN, Theory Division} 
\centerline{\it CH-1211 Geneva 23, Switzerland}
\vskip 1.2 true cm
\centerline{\bf Contents}
\vskip2ex

\vskip 0.8ex\noindent\kern4.0em
1.~Introduction

\vskip 0.8ex\noindent\kern4.0em
2.~Chiral gauge theories \& the gauge anomaly

\vskip 0.8ex\noindent\kern4.0em
3.~The regularization problem

\vskip 0.8ex\noindent\kern4.0em
4.~Weyl fermions from 4+1 dimensions

\vskip 0.8ex\noindent\kern4.0em
5.~The Ginsparg--Wilson relation

\vskip 0.8ex\noindent\kern4.0em
6.~Gauge-invariant lattice regularization of anomaly-free theories

\vskip-2ex

\section 1. Introduction

A characteristic feature of the electroweak interactions is that
the left- and right-handed components of the fermion
fields do not couple to the gauge fields in the same way.
The term chiral gauge theory is reserved for field theories
of this type, while all other gauge theories (such as QCD)
are referred to as vector-like, since the gauge fields only
couple to vector currents in this case.
At first sight the difference appears to be 
mathematically insignificant,
but it turns out that in many respects
chiral gauge theories are much more complicated.
Their definition beyond the
classical level, for example, is already highly non-trivial and 
it is in general extremely difficult to obtain any solid information
about their non-perturbative properties.

\subsection 1.1 Anomalies

Most of the peculiarities in chiral gauge theories
are related to the fact 
that the gauge symmetry tends to be violated by quantum effects.
Whether such anomalies
occur or not depends on the gauge group and the 
fermion multiplet. If they do,
the theory pro\-bably ceases to be meaningful, since
the gauge degrees of freedom are then no longer guaranteed to 
decouple from the physical modes and unitarity in the 
physical sector will consequently be lost.

In perturbation theory anomalies can arise
from fermion loops with three or more external legs such as the 
triangle subdiagram in fig.~1.
The origin of these anomalies and their topological significance
have been completely clarified in the eighties 
(see refs.~[\ref{Luis},\ref{Bertlmann}] for comprehensive reviews).
Moreover the rigorous work on algebraic renormalization 
[\ref{BRS}--\ref{PiguetSorella}] showed that
there are no further gauge anomalies in perturbation theory,
i.e.~the gauge symmetry can be preserved to all orders of the expansion
if the non-invariant terms cancel at one-loop order.

\topinsert
\vbox{
\vskip0.3cm
 
\centerline{\epsfxsize=6.0cm%
\epsfbox{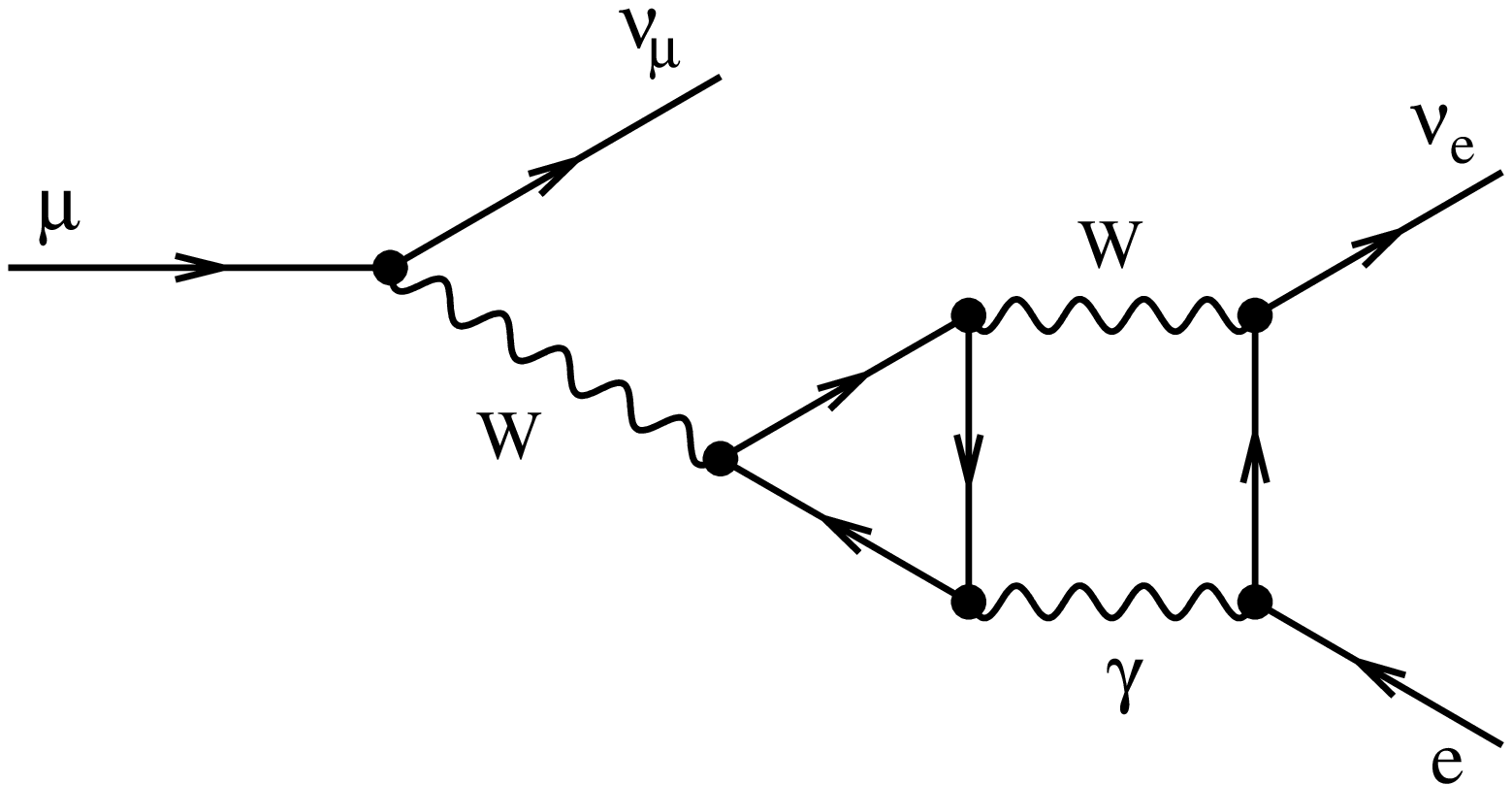}
}
\vskip0.3cm
\figurecaption{%
Feynman diagram contributing to the muon decay at two-loop order of 
the electroweak interactions.
The triangular subdiagram 
in this example is potentially anomalous and must be treated
with care to ensure that gauge invariance is preserved.
}
\vskip0.0cm
}
\endinsert

Additional constraints on
the gauge group and the fermion multiplet may
have to be imposed to ensure the consistency
of the theory also at the non-perturbative level.
The SU(2) gauge theory with a single left-handed fermion
in the fundamental representation, for example,
has no anomaly in perturbation theory,
but there is a non-perturbative 
{\it global anomaly}\/ that ruins the theory
[\ref{Witten}--\ref{BaerCampos}].
While this may be a rather special case, the remark shows
how delicate chiral gauge theories are and that there can be
unexpected complications beyond perturbation theory.

\subsection 1.2 Fundamental issues in chiral gauge theories

In the electroweak theory the perturbative anomalies cancel and 
the perturbation expansion
is thus well-defined and consistent to all orders
of the gauge couplings.
It is, however, hard to overlook the fact that the electroweak theory and 
chiral gauge theories in general appear to be rather artificial 
at the quantum level. 
Whether there is a deeper reason for this is unclear,
but a possible explanation could be
that chiral gauge theories are merely
low-energy effective descriptions of another
structure whose mathematical consistency does not depend on accidental
cancellations.
In such a framework it is conceivable that only the anomaly-free
fermion multiplets can decouple from the high-energy degrees of
freedom and the electroweak theory would then look a lot more natural
than is the case at present.

A second (but not totally unrelated) 
question is whether chiral gauge theories can be
regularized without giving up gauge invariance or other vital
properties such as the locality of the theory.
In a purely technical sense, such a regularization would
provide an example of a well-defined structure
that reduces to the desired chiral gauge theory at low energies
(i.e.~below the cutoff scale).
It might also be of some practical importance, since in the absence of
a gauge-invariant regularization many non-invariant counterterms must
be included in the lagrangian to
restore the gauge symmetry after renormalization and removal of the
cutoff [\ref{BRS}--\ref{PiguetSorella},\ref{Rome}--\ref{RomeReviewII}].
As a result the proof of the renormalizability of the theory 
and the computation of higher-order ra\-di\-ative corrections
are far more complicated than in vector-like theories
(see ref.~[\ref{GrassiHurth}], for example, and references quoted there).

Beyond perturbation theory, chiral gauge theories are still a largely
uncharted territory, and their formulation at this level is in fact
already a difficult task.
Vector-like gauge theories are more accessible in this respect, because
contrary to chiral theories 
they can easily be put on the lattice, and 
numerical simulations then provide a powerful tool to determine 
their properties.
Other approaches, such as the semi-classical appro\-xi\-mation, 
are helpful in
studying certain non-perturbative effects but cannot provide a
mathe\-ma\-ti\-cally solid definition of the theory beyond their range of
applicability.

\subsection 1.3 Scope of the lectures

Over the last few years significant progress has been made 
in all areas mentioned above
[\ref{PerfectDiracOperator}--\ref{SuzukiRealRep}].
It is now possible, for example, to put 
anomaly-free chiral gauge theories on the lattice,
to all orders of perturbation theory,
so that the gauge symmetry is exactly preserved
and without having to compromise in any other way
[\ref{SuzukiBRS},\ref{RegChGT}].
Some of these advances grew out of 
seemingly unrelated lines of research,
but most of them build
on earlier work on chiral gauge theories, such as the descent
equations [\ref{StoraI}--\ref{BaulieuII}] and the observation
that massive Dirac fermions in 4+1 dimensions
reduce to chiral fermions in 4 dimensions
under certain conditions
[\ref{RubakovShaposhnikov}--\ref{KikukawaAoyama}].

The aim in these lectures is to describe 
in simple terms some of the key elements of these
developments.
Since the lectures are intended for non-experts,
we shall start with a brief exposition of what everybody should know
about chiral gauge theories and the gauge anomaly 
in perturbation theory. The regularization problem is then 
described in more detail and a mini-introduction to lattice gauge
theory is included to set up basic notations.
Some familiarity with this subject is surely helpful but
will not be required. 

In the central part of the lectures we first discuss
how to obtain chiral fermions in 4 dimensions 
from Dirac fermions in 4+1 dimensions. This provides
an example of a natural mechanism for chiral fermions to arise,
and in a few lines it also leads us to the now famous
Ginsparg--Wilson relation [\ref{GinspargWilson}].
Very briefly this identity represents a new form of chiral
symmetry that can coexist with a momentum cutoff.
In particular, on a space-time lattice it
can be taken as the starting point for a general construction
of chiral lattice gauge theories, which is the last 
and most advanced topic 
that will be addressed in these lectures.

Chiral gauge theories are an old subject,
and it is clearly impossible to do justice to all the 
important contributions that have been made.
An excellent introduction to much of the earlier work 
is provided by the books of Bertlmann on anomalies [\ref{Bertlmann}] and
of Piguet and Sorella on algebraic renormalization [\ref{PiguetSorella}].
As far as the development of chiral lattice gauge theories goes,
probably the best source of information are the reviews
at the yearly lattice conferences 
[\ref{ShamirReview}--\ref{GoltermanReview}]. Many more references
can be found there and also a description of some of the 
alternative approaches to the problem.

\subsection Acknowledgements

I am indebted to Raymond Stora and Tobias Hurth for discussions on 
the renor\-ma\-li\-zation of chiral gauge theories and many useful pointers 
to the literature. Thanks also go to Fred Jeger\-leh\-ner for sharing 
his insights on dimensional regularization and to 
Yoshio Kikukawa and Hiroshi Suzuki with whom I had many 
enlightening con\-ver\-sations on the topics covered in this course.
The hospitality at Erice is legendary, and I would like 
to thank Gerard 't~Hooft, Gabriele Veneziano 
and Antonino Zi\-chi\-chi for the 
opportunity to lecture at this unique place.

\section 2. Chiral gauge theories \& the gauge anomaly

\vskip-2.5ex

\subsection 2.1 Classical theory

To avoid inessential complications, and since this 
is common practice in the more mathematical literature on the subject, 
the theory will be set up in euclidean space. 
The conventions for the Dirac matrices are
\equation{
  \{\dirac{\mu},\dirac{\nu}\}=2\delta_{\mu\nu},
  \qquad
  (\dirac{\mu})^{\dagger}=\dirac{\mu},
  \qquad
  \dirac{5}=\dirac{0}\dirac{1}\dirac{2}\dirac{3},
  \enum
}
and repeated indices are summed over unless stated otherwise.

For simplicity, we shall consider chiral gauge theories
with left-handed fermions only and no Higgs fields.
If we define the chiral projectors
$P_{\pm}=\frac{1}{2}(1\pm\dirac{5})$,
the fermion and antifermion fields thus satisfy the constraints
\equation{
  P_{-}\psi(x)=\psi(x),
  \qquad
  \psibar(x)P_{+}=\psibar(x).
  \enum
}
Under gauge transformations $\Lambda(x)$,
they transform according to 
\equation{
  \psi(x)\to R[\Lambda(x)]\psi(x),
  \qquad
  \psibar(x)\to \psibar(x)R[\Lambda(x)]^{-1},
  \enum
}
where $R$ is some unitary 
representation of the gauge group $G$. 
The associated gauge-covariant derivatives 
\equation{
  D_{\mu}=\partial_{\mu}+A_{\mu}^a(x)R(T^a)
  \enum
}
involve the group generators in the representation $R$
and the components $A_{\mu}^a(x)$ of the gauge field.
The notations here are such that the latter are real,
while the group generators are taken to be anti-hermitian.

With these definitions the euclidean action of the theory assumes
the form
\equation{
  S[A,\psibar,\psi]=
  \int\rmd^4x\left\{{1\over4g_0^2}\,F^a_{\mu\nu}(x)F^a_{\mu\nu}(x)+
  \psibar(x)\dirac{\mu}D_{\mu}\psi(x)\right\},
  \enum
}
where $F^a_{\mu\nu}(x)$ denotes the gauge field tensor and 
$g_0$ the gauge coupling. 
Although we agreed to consider only left-handed fermions,
it is now easy to check that charge conjugation 
maps left- to right-handed fermions in
the complex conjugate representation of the gauge group.
So if we set
\equation{
  R=R_{\rm left}\oplus \left(R_{\rm right}\right)^{\ast},
  \enum
}
the theory defined above is in fact equivalent to one with both left-
and right-handed fermions that transform according to the 
representations $R_{\rm left}$ and $R_{\rm right}$ respectively.

\subsection 2.2 Gauge anomaly

From the action (2.5) the Feynman rules can be 
deduced in the usual way. Gauge-fixing is 
required for this but does not need to be made explicit here,
since we shall only consider diagrams 
without internal gauge field lines.
Compared to QCD with massless quarks,
an important difference
is that the fermion propagator
\equation{
  \langle \psi(x)\psibar(y)\rangle_{g_0=0}=
  -i\int{\rmd^4p\over(2\pi)^4}\,\rme^{ip(x-y)}
  \,{\dirac{\mu}p_{\mu}\over p^2}
  P_{+}
  \enum
}
involves a chiral projector. 
Apart from this and the fact that the fermion-gauge-field vertices
are proportional to 
$R(T^a)$ instead of the group generator in the quark representation,
the Feynman rules are precisely the same.

We now consider the set of fermion one-loop diagrams 
with $n$ amputated external gauge field lines (see fig.~2).
In momentum space the external lines carry
in-going momenta $p_1,\ldots,p_n$, Lorentz indices $\mu_1,\ldots,\mu_n$ and
gauge group indices $a_1,\ldots,a_n$.
The sum of these diagrams is thus some function 
$V^{(n)}(p_1,\ldots,p_n)$ with these many indices.
It is not difficult to prove 
that the associated effective action
\equation{
  \Seff[A]=-\sum_{n=0}^{\infty}
  {1\over n!}\int{\rmd^4p_1\over(2\pi)^4}\ldots{\rmd^4p_n\over(2\pi)^4}\,
  (2\pi)^4\delta(p_1+\ldots+p_n)
  \noenum
  \next{2.0ex}
  \kern9.0em\times V^{(n)}(p_1,\ldots,p_n)^{a_1\ldots a_n}_{\mu_1\ldots \mu_n}
  \widetilde{A}^{a_1}_{\mu_1}(p_1)\ldots\widetilde{A}^{a_n}_{\mu_n}(p_n)
  \enum
}
(in which the gauge field plays the r\^ole of a classical source field)
is given by the functional integral
\equation{
  \rme^{-\Seff[A]}=\int\rmD[\psi]_{\rm left}\rmD[\psibar]_{\rm left}
  \exp\left\{
  -\int\rmd^4x\,\psibar(x)\dirac{\mu}D_{\mu}\psi(x)\right\}
  \enum
}
over the space of all left-handed fermion and antifermion fields.
Indeed, by expanding the integrand on the right-hand side of this
equation in powers of the gauge potential,
and applying Wick's rule to evaluate the gaussian integral,
all products of fermion one-loop diagrams are generated with the 
proper statistical factors to match the expansion of 
the left-hand side.

\topinsert
\vbox{
\vskip0.3cm
 
\centerline{\epsfxsize=4.0cm%
\epsfbox{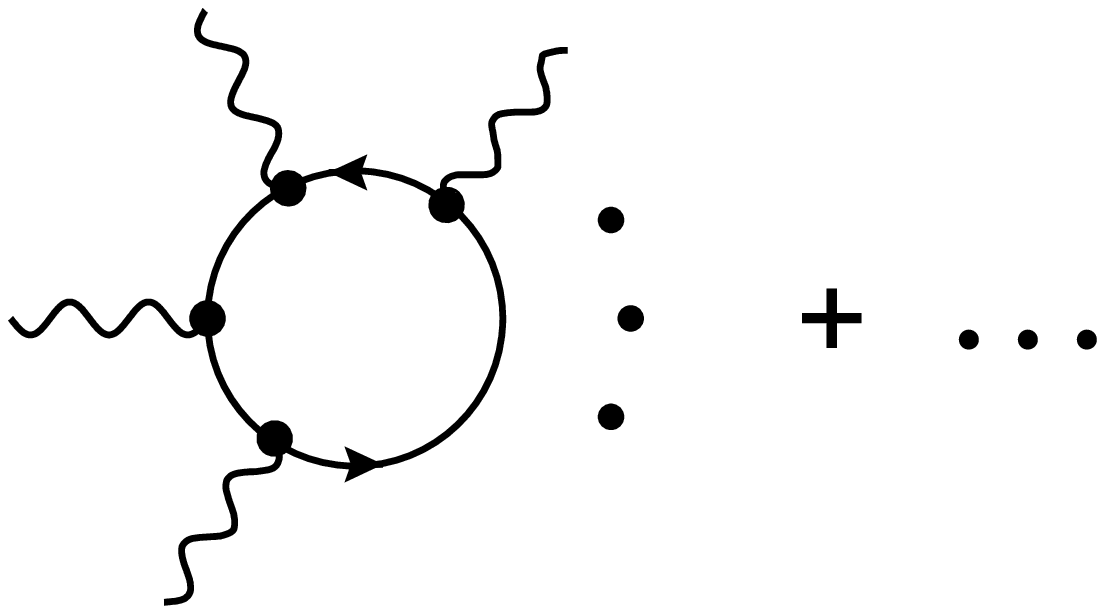}
}
\vskip0.3cm
\figurecaption{%
The vertices $V^{(n)}(p_1,\ldots,p_n)$ 
are equal to the sum of all $(n-1)!$
fermion one-loop diagrams with $n$ amputated external gauge field lines.
}
\vskip0.0cm
}
\endinsert

Since all entries in the functional integral respect the 
gauge symmetry, it seems obvious that the effective action
must be a gauge-invariant expression in the gauge potential.
However,
the fermion one-loop diagrams with less than 5 external lines
are ultra-violet-divergent and thus only incompletely defined. 
Now it can happen that any consistent way to make the diagrams
finite necessarily breaks the gauge symmetry of the effective action.
In this case we say that there is an anomaly.

The traditional way to cope with ultra-violet divergences is to introduce
a regula\-rization and to subtract the divergent terms from the diagrams
according to some renormalization scheme.
Different regularization and subtraction
prescriptions may give different results,
but power counting implies that the difference must be a
poly\-nom\-ial in the external momenta of a certain degree.
This amounts to adding a local term
\equation{
  \Delta\Seff[A]=\int\rmd^4x\,\Omega(x)
  \enum
}
to the effective action,
where $\Omega(x)$ is a polynomial in the gauge potential
$A^a_{\mu}(x)$ and its derivatives of dimension 4 or less.
In other words, 
the effective action in any scheme 
is the sum of the result obtained with
a particular prescription plus such a local term.

At this point an explicit calculation is required to determine
the effective action and its gauge transformation properties.
The diagrams may be worked out using a Pauli--Villars cutoff,
for example, which is defined by substituting
\equation{
  {1\over p^2}\to{1\over p^2}-{1\over p^2+\Lambda^2}
  \enum
}
in the fermion propagator (2.7) and taking the cutoff mass 
$\Lambda$ to infinity.
A more elegant computation starts from the observation that 
the functional integral (2.9) is proportional to the determinant of the 
Dirac operator $D=\dirac{\mu}D_{\mu}$.
The first-order variation of the effective action with respect
to the gauge field is thus given by %
\footnote\dag{\footnotefont%
Here and below the symbol ``Tr" denotes a trace
in field space, while the lower case ``tr" implies a trace over
Dirac, colour or flavour indices (depending on the context).}
\equation{
  \delta\Seff=-\Tr\bigl\{\delta D D^{-1}P_{+}\bigr\}
  =-\lim_{\epsilon\to0}\,
  \int_{\epsilon}^{\infty}\rmd t\,\Tr\bigl\{
  \delta DD^{\dagger}\rme^{-tDD^{\dagger}}P_{+}\bigr\}.
  \enum
}
For $\epsilon>0$ the integral on the right-hand side of this equation
is finite and provides a regularization of $\delta\Seff$.
Differential geometric methods (the heat kernel expansion)
may then be applied to study the limit $\epsilon\to0$ and
to determine the finite part of the effective action in this scheme
[\ref{Leutwyler}].

Once this is achieved, we may ask whether the now well-defined
effective action is invariant under gauge variations 
\equation{
  \delta A^a_{\mu}(x)=\partial_{\mu}\omega^a(x)
  +f^{abc}A_{\mu}^b(x)\omega^c(x)
  \enum
}
of the gauge potential (where $\omega^a(x)$ denotes an infinitesimal
gauge transformation and 
$f^{abc}$ the structure constants of the gauge group). 
As it turns out, the effective action is not invariant in general,
but its transformation behaviour can be worked out explicitly 
and is given by
\equation{
  \delta\Seff=
  {i\over192\pi^2}\int\rmd^4x\,
  \epsilon_{\mu\nu\rho\sigma}d_R^{abc}
  \omega^a(x)
  \Bigl\{\partial_{\mu}A^b_{\nu}(x)\partial_{\rho}A^c_{\sigma}(x)
  +\frac{1}{2}\partial_{\mu}\bigl[A^b_{\nu}(x)F^c_{\rho\sigma}(x)
  \bigr]\Bigr\}
  \noenum
  \next{2.0ex}
  {\phantom{\delta\Seff=}}+\int\rmd^4x\,\delta\Omega(x),
  \enum
  \next{2.0ex}
  d_R^{abc}=
  2i\,\tr\bigl\{R(T^a)R(T^b)R(T^c)+(b\leftrightarrow c)\bigr\}
  \enum
}
(see ref.~[\ref{Leutwyler}], for example).
Following our discussion above,
the variation of an ar\-bi\-trary local term has been included 
on the right-hand side of eq.~(2.14), and the formula thus holds
for any regularization and subtraction scheme.

The algebraic structure of the two terms in eq.~(2.14)
is such that they cannot cancel each other.
Unless $d_R^{abc}$ vanishes identically,
gauge invariance is hence violated at one-loop order 
of perturbation theory. The unphysical longitudinal 
components of the gauge field must then be expected
to couple to the fermions in higher-loop diagrams such as the 
one shown in fig.~1. As a consequence unitarity in the physical
sector is lost and the theory becomes unusable.

\subsection 2.3 Anomaly-free fermion representations

The second term in eq.~(2.14) can always be cancelled
by adding a local counterterm to the gauge-field action,
and if $d_R^{abc}$ happens to be equal to zero,
the anomaly disappears and the gauge symmetry is preserved.
Fermion representations with vanishing $d$-symbol are thus referred to 
as {\it anomaly-free}. In a U(1) theory, for example,
there is only one generator, and up to unitary 
transformations all representations are of the form
\equation{
  R(T^1)=i\times\hbox{diag}(\rme_1,\ldots,\rme_N),
  \qquad
  d^{111}_R=4\sum_{\alpha=1}^N\rme_{\alpha}^3,
  \enum
}
where $N$ denotes the number of fermions and 
$\rme_{\alpha}$ their charges. The anomaly-free representations
are then precisely those where the sum of the cubes of the charges vanishes.

Whether a given representation $R$ is anomaly-free or not can often 
be quickly de\-ci\-ded. 
First note that $d_R^{abc}$ is real,
since the symmetrized product of the 
generators in the trace (2.15) is anti-hermitian.
Real and pseudo-real representations are
hence always anomaly-free. In particular, chiral theories 
with gauge group SU(2) are safe from anomalies,
because SU(2) has only such representations.

The groups SU($n$) with $n\geq3$, on the other hand, have
complex representations and at least some of them are not anomaly-free.
For any representation $R$ of these groups, we have
$d_R^{abc}=c_Rd^{abc}$
where $d^{abc}$ denotes the $d$-symbol in the fundamental
representation. 
The representation is thus anomaly-free if and only if
\equation{
  c_R=\tr\bigl\{R(T)^3\bigr\}\kern-1pt\bigm/\kern-1pt\tr\bigl\{T^3\bigr\},
  \qquad
  T\equiv i\times\hbox{diag}(1,\ldots,1,1-n),
  \enum
}
vanishes. In the standard SU(5) grand unified theory
[\ref{SUfive}], for example, the fermions transform according
to the representation
${\bf 5}^{\ast}\oplus{\bf 10}$, which has 
$c_R=c_{\kern1pt{\bf 5}^{\ast}}+c_{\kern1pt{\bf 10}}=0$.

At the level of the Lie algebra, a general compact Lie group
decomposes into a product of U(1) factors and simple groups.
Anomaly-free representations $R$ of such a group must obviously reduce
to anomaly-free representations of all its factors.
Going through Cartan's list of the simple Lie algebras,
the only ones that admit totally symmetric invariant tensors of rank 3
are those associated with SU($n$), $n\geq3$~[\ref{Zelobenko}]. 
The representations of all other simple groups are thus 
anomaly-free, i.e.~only the factors with the Lie algebra
of SU($n$) and the abelian
factors need to be checked.

Even if the representation is anomaly-free when reduced to 
any one of the factors of the group, {\it mixed anomalies}
can still be present,
where some components of $d_R^{abc}$ with indices 
belonging to different factors do not vanish.
In particular, the components
\equation{
  d_R^{abc}=4i\,
  \tr\bigl\{\kern1pt\underbrace{R(T^a)R(T^b)}_{\rm simple\;factor}
  \kern0pt\underbrace{R(T^c)}_{\rm U(1)\;factor}
  \kern-5pt\bigr\}
  \enum
}
need not be equal to zero and there can also be mixed anomalies
between different U(1) factors (but not between different simple factors).

\section 3. The regularization problem

Vector-like theories can be regularized without breaking the 
gauge symmetry, using dimensional regularization, for example, or 
by putting them on a lattice. 
It is clear from the beginning that the situation in the chiral case
has to be more com\-plicated, because gauge-invariant regularizations
surely can only exist if the fermion multiplet is anomaly-free.
In particular,
{\it any consistent regularization that preserves the gauge symmetry
must refer to the fermion representation $R$}.

This simple observation alone implies that none of
the widely used regularization schemes can be expected to 
provide a solution of the problem.
The difficulty shows up in various ways,
but the no-go theorem was always confirmed and 
at some point the conclusion was drawn that chiral gauge
theories cannot be regularized without breaking the gauge symmetry.
In the anomaly-free case, the best one can hope for is then
that the symmetry will be restored after renormalization and removal
of the regularization
[\ref{BRS}--\ref{PiguetSorella},\ref{Rome}--\ref{RomeReviewII}].

Although a straightforward application of the standard 
methods does not lead to a gauge-invariant regularization
of chiral theories,
it is instructive to have a closer look at some of them
and to determine what exactly goes wrong.
A significant part of the present section is
devoted to the lattice regularization, which 
is a particularly well studied case in this respect.

\subsection 3.1 Naive dimensional regularization

Probably the most economical regularization method that we know of is 
dimensional regularization [\ref{DimRegI}]. To be able to apply it to 
chiral gauge theories, the Dirac algebra has to be extended to $d$
dimensions and an unambiguous definition of $\dirac{5}$ must be supplied.
The so-called naive dimensional regularization scheme is characterized by
\equation{
  \{\dirac{\mu},\dirac{\nu}\}=2\delta_{\mu\nu},
  \qquad
  \{\dirac{5},\dirac{\mu}\}=0,
  \qquad
  (\dirac{5})^2=1, \enum
  \next{2.5ex}
  \delta_{\mu\mu}=d,
  \qquad
  \tr\{1\}=4, \enum
}
plus the usual integration rules in $d$ dimensions. 
This scheme is known to be algebraically consistent (any Feynman diagram with
non-exceptional external momenta is assigned a well-defined
meromorphic function of $d$) and it also preserves the gauge symmetry
of the theory.

The prescription nevertheless fails to provide an acceptable
regularization of chiral gauge theories, because 
the rules (3.1),(3.2) together with the requirement of analyticity 
in $d$ imply
\equation{
  \tr\{\dirac{5}\dirac{\mu_1}\ldots\dirac{\mu_{2n}}\}=0
  \enum
}
for all $n$.
It is not difficult to establish this result (appendix A).
To fully understand its significance,  
let us again consider the effective action that we discussed
in the previous section. 

Fermion one-loop diagrams with $n$ external gauge field lines
are proportional to the trace of a product of $2n$ Dirac matrices
and $n$ chiral projectors. Since $\dirac{5}$ anti-commutes
with the Dirac matrices,
the latter can be combined to a single projector
and the trace then assumes the form
$\tr\{P_{+}\dirac{\mu_1}\ldots\dirac{\mu_{2n}}\}$.
Equation (3.3) says that
naive dimensional regularization sets the parity-odd part 
of the trace to zero. The anomaly is thus avoided, but at the same
time we know that the answer is incorrect, because a direct 
evaluation of the trace in four dimensions
(using ordinary Dirac matrices) yields a sum of
parity-odd terms proportional to
$\epsilon_{\mu\nu\rho\sigma}$ if $n\geq2$.

\subsection 3.2 Regularization with higher-derivative terms

In general the inclusion of higher-derivative terms 
in the action has a regularizing effect,
because the extra derivatives lead to propagators that 
decrease more rapidly at large momenta 
[\ref{SlavnovI}--\ref{FaddeevSlavnov}].
The transverse part of the gauge field propagator, for example,
decays like $(p^2)^{-2}$ at momenta $p$ much greater than $\Lambda$ 
if the term 
\equation{
  {1\over\Lambda^2}
  \int\rmd^4x\,D_{\mu}F_{\nu\rho}^a(x)D_{\mu}F_{\nu\rho}^a(x)
  \enum
}
is added to the action. The mass $\Lambda$ then plays the r\^ole 
of an ultra-violet cutoff that
is to be sent to infinity at the end of the calculation.

In this strictly four-dimensional approach,
the definition of the Dirac matrices is not an issue,
but the question of whether the gauge symmetry can be preserved
needs to be carefully discussed (even in the pure gauge theory).
First note that higher-derivative terms have to be included to 
suppress the propagators of the ghost fields and 
the longitudinal components of the gauge field at large momenta
after fixing the gauge.
This can be done so that the total gauge-fixed action is invariant under 
the Becchi--Rouet--Stora (BRS) transformation [\ref{BRS}--\ref{PiguetSorella}],
and the gauge symmetry is then guaranteed to be preserved
at the level of the physical amplitudes.

Power-counting now shows, however, that no matter which higher-derivative
terms are added, some one-loop diagrams always remain unregularized, because
terms like (3.4) not only modify the 
propagators but also give rise to additional vertices.
These vertices are required to ensure the gauge invariance of the regularized
theory, and while they are suppressed by inverse powers of 
the cutoff $\Lambda$, their insertion
increases the degree of divergence of the diagrams.
As a result some of these diagrams have positive degree.

Further regularization prescriptions are thus needed
to make all diagrams finite. 
In refs.~[\ref{SlavnovI}--\ref{FaddeevSlavnov}],
for example, it has been proposed 
to include a set of Pauli--Villars ghost fields with
properly chosen couplings to the gauge field.
Such hybrid regula\-ri\-zations are delicate, and it is
not easy to prove the correctness of the procedure.
Some of these schemes are in fact known to yield wrong results
[\ref{MartinEtAlI},\ref{MartinEtAlII}], while others require
intermediate dimensional regularization [\ref{AsoreyFalceto}] 
or include non-local terms in the action [\ref{BakeyevSlavnov}].
Adding higher-derivative terms is, therefore, still
not a very transparent regularization method, not even 
in vector-like theories
(see, however, ref.~[\ref{DeminovSlavnov}]).

\subsection 3.3 Lattice gauge theory

The lattice formulation of quantum field theories 
provides a good starting point for
non-perturbative studies, 
and this is no doubt the principal reason
why lattice field theory continues to be a popular research topic
for now more than 25 years.
In the present context 
we are interested in perturbation theory, where the 
lattice makes all diagrams manifestly finite.

Lattice field theories are usually set up 
on a hypercubic lattice with spacing $a$ (see fig.~3). 
A fermion field $\psi(x)$, for example, is then 
simply an assignment of a Dirac spinor to each lattice point
\equation{
  x=a\left(n_0,n_1,n_2,n_3\right),
  \qquad
  n_{\mu}\in\gz.
  \enum
}
The Fourier representation of any such field,
\equation{
  \psi(x)=\int_{-\pi/a}^{\pi/a}{\rmd^4p\over(2\pi)^4}\,
  \rme^{ipx}\kern1pt\widetilde{\psi}(p),
  \enum
}
involves an integration over momenta in a bounded region only 
(the Brillouin zone),
and lattice field theories thus have a built-in momentum cutoff
of order $1/a$.

\topinsert
\vbox{
\vskip0.3cm
 
\hskip2.5cm\epsfxsize=3.5cm\epsfbox{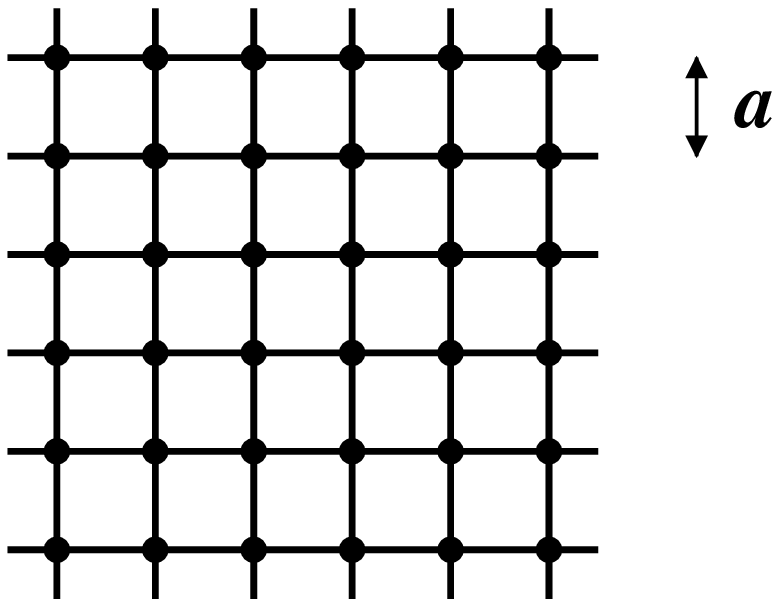}

\vskip-1.8cm
\hskip7.5cm\epsfxsize=2.3cm\epsfbox{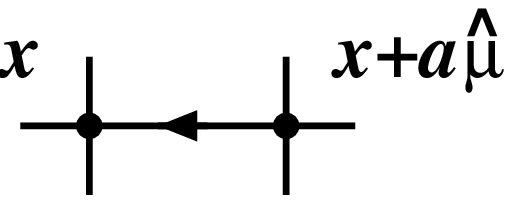}

\vskip1.3cm
\figurecaption{%
In lattice gauge theory the fermion fields $\psi(x)$ 
reside on the points $x$ of a regular hypercubic lattice 
with spacing $a$ and the gauge field variables $U(x,\mu)$ 
on the directed links $(x,x+a\hat{\mu})$ of 
the lattice (expanded view on the right).
}
\vskip0.0cm
}
\endinsert

On the lattice the forward and backward difference operators
\equation{
  \drv{\mu}\psi(x)=\{\psi(x+a\hat{\mu})-\psi(x)\}/a,
  \enum
  \next{2.5ex}
  \drvstar{\mu}\psi(x)=\{\psi(x)-\psi(x-a\hat{\mu})\}/a,
  \enum
}
(where $\hat{\mu}$ denotes the unit vector in direction $\mu$)
can be taken as substitutes for 
the partial differential operators in the 
continuum theory. 
It is useful to introduce them both, because $\drvstar{\mu}$ is equal 
to minus the adjoint of $\drv{\mu}$ and vice versa. The lattice laplacian,
for example, can then be compactly written as $\drvstar{\mu}\drv{\mu}$.

Lattice Dirac operators that reduce to the continuum Dirac operator 
in the limit $a\to0$ are now also easily constructed.
Partly because of its simplicity,
the expression proposed by Wilson in 1974 [\ref{Wilson}], 
\equation{
  D_{\rm w}=\frac{1}{2}\left\{\dirac{\mu}(\drvstar{\mu}+\drv{\mu})
  -a\drvstar{\mu}\drv{\mu}\right\},
  \enum
}
is still widely used today, but there are many other
acceptable lattice Dirac operators that all 
lead to the same continuum limit (the significance of the second term
in the definition of $D_{\rm w}$ will be clarified in the next
subsection).

In the continuum theory, gauge-covariant differentiation 
requires the introduction of gauge potentials with the appropriate 
transformation behaviour.
Lattice gauge fields serve exactly the same purpose, except
that here we are dealing with
difference instead of differential operators.
Explicitly, if we assume that 
the gauge group acts on lattice fermion fields in the obvious
way [eq.~(2.3)], and if 
$U(x,\mu)$ is a lattice field
with values in the gauge group, which transforms according to
\equation{
  U(x,\mu)\to\Lambda(x)U(x,\mu)\Lambda(x+a\hat{\mu})^{-1},
  \enum
}
it is trivial to check that the difference operators
\equation{
  \nab{\mu}\psi(x)=\{
  R[U(x,\mu)]\psi(x+a\hat{\mu})-\psi(x)\}/a,
  \enum
  \next{2.5ex}
  \nabstar{\mu}\psi(x)=\{
  \psi(x)-R[U(x-a\hat{\mu},\mu)]^{-1}\psi(x-a\hat{\mu})\}/a,
  \enum
}
are gauge-covariant (cf.~fig.~3). 
On the lattice, gauge fields are thus 
represented by group-valued fields $U(x,\mu)$ rather than 
vector fields with values in the Lie algebra of the gauge group. 
This seems a bit strange at first sight, but 
should not be given too much weight,
because the difference is mainly a matter of notation.
In perturbation theory, for example, the parametrization
\equation{
  U(x,\mu)=\exp\bigl\{aA^a_{\mu}(x)T^a\bigr\}=1+aA^a_{\mu}(x)T^a+\ldots
  \enum
}
is usually employed, which
leads again to a description in terms of
a gauge potential $A^a_{\mu}(x)$. The covariant
difference operators $\nab{\mu}$ and $\nabstar{\mu}$ are then also
easily seen to converge to the covariant differential operator
$D_{\mu}$ in the limit $a\to0$.

It is now obvious that the lattice fermion action
\equation{
  S_{\rm F}[U,\psibar,\psi]=a^4\sum_x\psibar(x)D_{\rm w}\psi(x)
  \enum
}
preserves the gauge symmetry if the ordinary difference operators
in the definition (3.9) of the Wilson--Dirac operator
are replaced by covariant ones.
Invariant gauge field actions are also not difficult to construct
[\ref{Wilson}--\ref{RotheBook}]
and the bottom line is then
that vector-like theories such as QCD can be 
put on the lattice without breaking the gauge symmetry or running
into any other fundamental difficulty.
The lattice formulation of these theories has in fact long 
been shown to provide
a completely consistent regularization to all orders of 
perturbation theory. In particular, gauge-fixing is
a rigorous procedure in this framework, and the existence
of the continuum limit has been established using
the Reisz power-counting theorem and the BRS symmetry
[\ref{LesHouches}--\ref{BFM}].

\subsection 3.4 The Nielsen--Ninomiya no-go theorem

At this point it seems that chiral lattice gauge theories 
can be obtained simply by imposing the constraints
(2.2) on the fermion and antifermion fields.
To understand why this is not so, it suffices to 
consider the free fermion theory. 
First note that
\equation{
  D_{\rm w}\kern1pt\rme^{ipx}u=
  \bigl\{i\dirac{\mu}\ring{p}_{\mu}+\frac{1}{2}a\hat{p}^2\bigr\}
  \kern1pt\rme^{ipx}u,
  \enum
  \next{2.0ex}
  \ring{p}_{\mu}=(1/a)\sin(ap_{\mu}),
  \qquad
  \hat{p}_{\mu}=(2/a)\sin(ap_{\mu}/2),
  \enum
}
for any four-momentum $p$ and Dirac spinor $u$.
The fermion propagator (which co\-in\-cides with  
the Green function of the Wilson--Dirac operator)
is thus given by
\equation{
  \langle\psi(x)\psibar(y)\rangle=-i
  \int_{-\pi/a}^{\pi/a}{\rmd^4p\over(2\pi)^4}\,
  \rme^{ip(x-y)}\kern1pt
  {\dirac{\mu}\ring{p}_{\mu}+i\frac{1}{2}a\hat{p}^2
  \over\ring{p}^2+\frac{1}{4}a^2(\hat{p}^2)^2}.
  \enum
}
In particular, the propagator has no singularities
in momentum space
other than the expected one-particle pole at $p=0$.

Now if we impose the chiral constraints (2.2), the part of the 
Wilson--Dirac opera\-tor that survives in the fermion action (3.14) is
\equation{
  P_{+}D_{\rm w}P_{-}=
  \frac{1}{2}P_{+}\dirac{\mu}(\drvstar{\mu}+\drv{\mu}).
  \enum
}
In momentum space the denominator of the chiral propagator 
is hence equal to $\ring{p}^2$ and not $\ring{p}^2+\frac{1}{4}a^2(\hat{p}^2)^2$
as in the Dirac case. As a consequence there are now poles 
at all momenta with components 
$p_{\mu}\in\{0,\pm\pi/a\}$, and these poles
effectively describe separate fermion species.
If the fermion field is coupled to an external gauge field, for example,
the effective action turns out to coincide, in the continuum limit, with
the result expected for a theory with these many fermion flavours.
 
The fermion doubling problem (as it is called in the literature)
is not specific to any particular lattice formulation.
The Nielsen--Ninomiya no-go theorem [\ref{NN},\ref{Friedan}]
in fact asserts that
{\it the chiral projection with the projectors $P_{\pm}$ always
leads to unphysical poles in the fermion propagator if the 
chosen lattice Dirac operator is local}. 
Having more fermions than intended
or giving up the locality of the theory 
is clearly unacceptable.
There were many attempts to circumvent this difficulty,
but it is only in the last two years that a solution has finally been found.
A particularly appealing approach to this solution 
starts from fermions in 4+1 dimensions,
which is the topic of the next section.

\appendix A. Proof of eq.~(3.3)

After inserting
the identity $\dirac{\mu}\dirac{\mu}=d$
and cycling one of the 
factors $\dirac{\mu}$ around the trace, using the anticommutation rules
(3.1), an equation of the form 
\equation{
  (d-2n)\kern1pt
  \tr\{\dirac{5}\dirac{\mu_1}\ldots\dirac{\mu_{2n}}\}+\ldots=0
  \enum
}
is obtained,
where the ellipses stand for a sum of traces with $2n-2$ Dirac matrices.
In particular, for $n=0$ the algebra yields $d\kern1pt\tr\{\dirac{5}\}=0$
and thus $\tr\{\dirac{5}\}=0$.
Equation (3.3) now follows from eq.~(A.1) by induction over $n$.

Note that we cannot escape the argument by
assigning a non-zero value to the traces at $d=2n$ only,
because in a dimensional regularization scheme the Feynman 
integrals must be analytic functions of $d$ and such
singular solutions of the recursion (A.1) are hence not acceptable.

\section 4. Weyl fermions from 4+1 dimensions 
            
The descent equations [\ref{StoraI}--\ref{BaulieuII}]
were probably the first instance where a connection between chiral fermions
and field theories in higher dimensions was made.
At the time this seemed to be a purely algebraic observation,
relating the Chern character 
in six dimensions via the Chern--Simons density in five dimensions
to the gauge anomaly.
The relationship in fact
goes far beyond the formal level
[\ref{RubakovShaposhnikov}--\ref{KikukawaAoyama}],
and while the discussion
in the present section is rather limited in scope, 
it can be regarded as a first step in this direction.

\subsection 4.1 Domain wall fermions

Since the fifth dimension is going to play a special r\^ole,
we do not employ a covariant notation, i.e.~the extra coordinate
will be denoted by $s$ and Lorentz indices 
$\mu,\nu,\ldots$ label the four physical directions as before.
In the presence of a scalar background field $\phi(s)$, the 
Dirac operator in 4+1 dimensions is then given by
\equation{
  D_5=\dirac{\mu}\partial_{\mu}+\dirac{5}\partial_s-\phi(s).
  \enum
}
The question of where the background field comes from will 
not concern us here, and we shall simply assume 
that it has the shape of 
a step function of height $M$ and width $1/M$ (see fig.~4). 
Such a background field defines a
domain wall separating the half-spaces $s<0$ and $s>0$
from each other.

\topinsert
\vbox{
\vskip0.3cm
 
\centerline{\epsfxsize=3.3cm%
\epsfbox{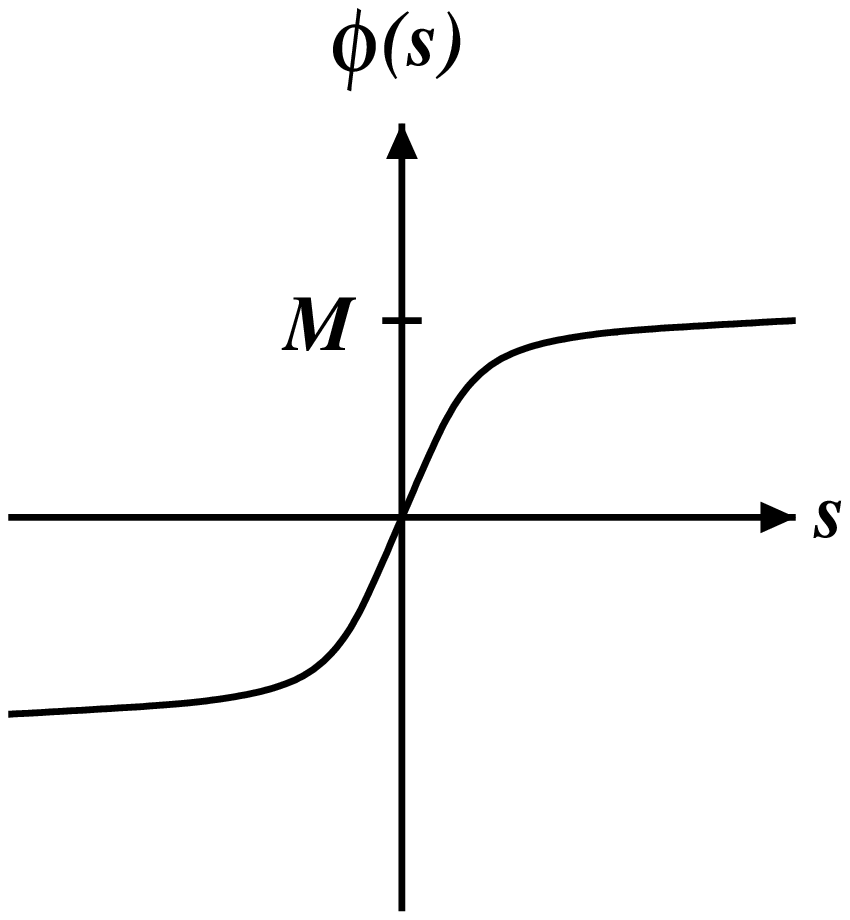}
}
\vskip0.3cm
\figurecaption{%
Qualitative shape of the background field $\phi(s)$.
The prototype of such a function is $\phi(s)=M\tanh(Ms)$, 
but its analytic form will not be needed here.
}
\vskip0.0cm
}
\endinsert

We now show that the domain wall affects  
the physical properties of the fermions
in an interesting way [\ref{RubakovShaposhnikov},\ref{CallanHarvey}].
Fermion fields in 4+1 dimensions 
with energy $E$ and momentum ${\bf p}$ in 4 dimensions 
are of the form
\equation{
  \chi(x,s)=\rme^{ipx}u(s),
  \qquad
  p=(iE,{\bf p})
  \enum
}
(recall that we are using a euclidean metric, where the energy
components of phy\-si\-cal four-momenta are purely imaginary).
The Dirac equation $D_5\chi(x,s)=0$ then becomes
\equation{
  \{\dirac{5}\partial_s-\phi(s)\}\kern1pt u(s)=
  -i\dirac{\mu}p_{\mu}\kern1pt u(s),
  \enum
}
and after multiplication from the left with $-i\dirac{\mu}p_{\mu}$,
this leads to the equation
\equation{
  \{-\partial_s^2+V(s)\}\kern1pt u(s)=m^2u(s),
  \enum
  \next{2.0ex}
  V(s)=\dirac{5}\partial_s\phi(s)+\phi(s)^2,
  \qquad 
  m^2=E^2-{\bf p}^2.
  \enum
}
The possible fermion masses $m$ are thus determined by the eigenvalues
of a certain differential operator. In other words,
from the point of view of the four-dimensional world, 
the presence of the extra dimension results in a tower 
of fermions with these masses.

The operator on the left-hand side of eq.~(4.4) commutes with $\dirac{5}$
and its eigenfunctions may hence be assumed to have definite 
chirality. We now distinguish three cases.

\vskip1ex
\noindent
(a)~{\sl Continuous spectrum}. Independently of the chirality,
the asymptotic value of the potential $V(s)$ at large $|s|$ is equal
to $M^2$, and the spectrum of the operator 
thus includes the half-line $m^2\geq M^2$ (see fig.~5).

\topinsert
\vbox{
\vskip0.3cm
 
\centerline{\epsfxsize=3.3cm%
\epsfbox{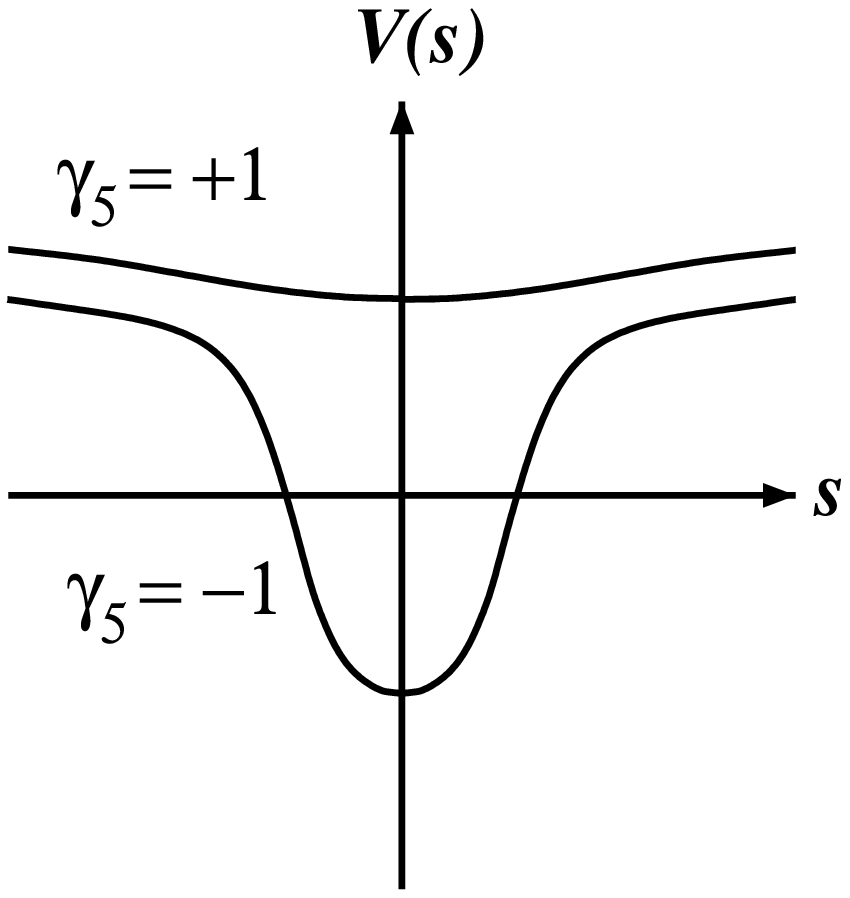}
}
\vskip0.3cm
\figurecaption{%
For negative chirality,
the potential $V(s)$ has a well at $s=0$ in which the massless mode
is trapped. The right-handed modes, on the other hand, are always heavy
independently of whether they are bound to the domain wall or not.
}
\vskip0.0cm
}
\endinsert

\vskip1ex
\noindent
(b)~{\sl Discrete spectrum}. Eigenfunctions with eigenvalues
below $M^2$ decay exponentially and this part of the 
mass spectrum is thus purely discrete.
Negative eigenvalues (tachyons) are excluded, since
\equation{
  -\partial_s^2+V(s)=
  \{-\dirac{5}\partial_s+\phi(s)\}^{\dagger}\kern1pt
  \{-\dirac{5}\partial_s+\phi(s)\}
  \enum
}
is a non-negative operator. Moreover, all non-zero mass values $m$ must
be of order $M$, since there is no other characteristic scale around.

\vskip1ex
\noindent
(c)~{\sl Massless modes}. In this case the Dirac equation reduces to
\equation{
    \{-\dirac{5}\partial_s+\phi(s)\}\kern1pt u(s)=0,
    \qquad\dirac{\mu}p_{\mu}\kern1pt u(s)=0, 
    \enum
}
from which we infer that
\equation{
  u(s)=\exp\left\{\pm\int_0^s\rmd t\,\phi(t)\right\}v,
  \qquad 
  \dirac{\mu}p_{\mu}\kern1pt v=0,
  \qquad
  P_{\pm}v=v.
  \enum
}
While the positive chirality solution
grows exponentially at large $|s|$, the left-handed mode is 
normalizable and $m^2=0$ thus belongs to the discrete mass spectrum.

\vskip1ex

To sum up we have found that all fermion modes, except for one, 
have mass $m$ of order $M$ or larger. The massless mode
is left-handed and its wave function 
[which is given by eqs.~(4.2),(4.8)]
falls off exponentially in the extra dimension.
At energies $E$ far below $M$, the theory thus describes 
a single left-handed chiral fermion that propagates along 
the domain wall. 
Effectively a dimensional reduction from 
4+1 to 4 dimensions is taking place, since it requires an energy
proportional to $M$ or larger to excite the higher fermion modes
[\ref{RubakovShaposhnikov},\ref{CallanHarvey}].

\subsection 4.2 Fermion propagator

An attractive feature of this mechanism is its stability against
various changes of the set-up. In particular, the precise form
of the shape function $\phi(s)$ is irrelevant, and we may in fact 
replace the domain wall through a boundary with Dirichlet
boundary conditions. The fermion propagator can then
be worked out explicitly and
provides further insight into what dimensional reduction means
in the present context.

So let us consider Dirac fields $\chi(x,s)$ in the half-space $s\geq0$
that satisfy
\equation{
  \left.P_{+}\chi(x,s)\right|_{s=0}=0.
  \enum
}
In this formulation 
a background field is no longer required, and for the 
Dirac opera\-tor we can simply take 
\equation{
  D_5=D_4+\dirac{5}\partial_s-M,
  \qquad
  D_4=\dirac{\mu}\partial_{\mu}.
  \enum
}
The spectrum of fermion masses is then 
as before, with a left-handed Weyl fermion that 
moves along the boundary at $s=0$.

In 4+1 dimensions the fermion propagator is defined by
\equation{
  \left.D_5G(x,s;y,t)\right|_{s,t>0}=\delta(x-y)\delta(s-t),
  \qquad
  \left.P_{+}G(x,s;y,t)\right|_{s=0}=0.
  \enum
}
Since $D_4$ does not depend on the extra coordinate,
it is straightforward to solve these 
equations (appendix). In particular, for $x\neq y$ 
the result 
\equation{
  \left.G(x,s;y,t)\right|_{s=t=0}=2M\times P_{-}S(x,y)P_{+}
  \enum
}
is obtained,
where $S(x,y)$ denotes the Green function of the operator
\equation{
  D=M+(D_4-M)\left[1-(D_4/M)^2\right]^{-1/2}
  \enum
}
(that acts on Dirac fields in four dimensions).
The inverse square root in this formula is defined in the obvious way,
taking into account the fact that the operator in square brackets 
is hermitian and bounded from below by $1$. 

The propagator (4.12) describes the fermion propagation
along the boundary of the five-dimensional world and should
reflect the presence of the massless mode. This is indeed
the case, since at four-momenta far below $M$, the operator $D$
is given by
\equation{
  D=D_4\left\{1-{D_4\over2M}+\ldots\right\},
  \enum
}
i.e.~it coincides with the four-dimensional Dirac operator 
up to terms of order $1/M$. In particular, the right-hand side of eq.~(4.12)
is equal to the propagator of a free left-handed fermion in this kinematical 
regime. The reduction from Dirac fermions in five dimensions to 
Weyl fermions in four dimensions thus also works out at the 
level of the underlying field theory.

\appendix B. Analytic formula for $G(x,s;y,t)$

To simplify the notation, 
we interpret the propagator at fixed $s$ and $t$ as an
operator ${\cal G}(s,t)$ 
that acts on Dirac fields $\psi(x)$ in four dimensions according to
\equation{
  {\cal G}(s,t)\psi(x)=\int\rmd^4y\,G(x,s;y,t)\psi(y).
  \enum
}  
In operator form, the equations that need to be solved are
\equation{
  \left.D_5{\cal G}(s,t)\right|_{s,t>0}=\delta(s-t),
  \qquad
  \left.P_{+}{\cal G}(s,t)\right|_{s=0}=0,
  \enum
}
and this can be done elegantly in terms of the hermitian operators
\equation{
  Q=\dirac{5}(M-D_4),
  \qquad
  \Phat_{\pm}=\frac{1}{2}\bigl\{1\pm Q\kern1pt[Q^2]^{-1/2}\bigr\}.
  \enum
}
The latter are just the projectors to the subspaces of eigenvectors of $Q$
with positive and negative eigenvalues. Note that the definition
of these subspaces is unambiguous, since 
all eigenvalues of $Q$ have absolute magnitude greater than or equal to $M$.

The claim is now that the operator
\equation{
  {\cal G}(s,t)=
  \left\{
  \rme^{(s-t)Q}\bigl[\theta(s-t)\Phat_{-}-\theta(t-s)\Phat_{+}\bigr]
  +\rme^{sQ}\Phat_{-}
  {2M\over D}\Phat_{+}\rme^{-tQ}
  \right\}\dirac{5}
  \enum
}
has all the required properties. One of these is that the expression
should go to zero if $|s-t|$ becomes large, which is the case since 
the exponentials are multiplied with the appropriate 
product of step functions and
projectors $\Phat_{\pm}$. Using $D_5=\dirac{5}(\partial_s-Q)$,
it is also trivial to show that the Dirac equation holds.
Finally, starting from the definitions (4.13) and (4.17), the identity
\equation{
  2MP_{+}\Phat_{-}=P_{+}D
  \enum
}
may be deduced, and this immediately implies that the operator (B.4)
satisfies the boundary condition.

\section 5. The Ginsparg--Wilson relation

\vskip-2.5ex

\subsection 5.1 Kaplan's observation

As already mentioned, the occurrence of 
a massless fermion mode in the presence of a domain wall 
is a generic effect. In 1992 Kaplan noted that the 
massless mode persists even if the five-dimensional space
is replaced by a lattice [\ref{Kaplan}].
It is natural to set $M=1/a$ in this context
(where~$a$ denotes the lattice spacing),
and the masses of the heavy modes are then of the order of the 
momentum cutoff or larger.

An inaccurate but often quoted form of 
the Nielsen--Ninomiya theorem states
that it is not possible to have a single Weyl fermion
on a four-dimensional lattice. 
This was, however, precisely what Kaplan obtained. 
In particular, he showed that there was no doubling of fermion species
in four dimensions if the Wilson--Dirac operator was used in five
dimensions (cf.~subsect.~3.3).

The resolution of this puzzle basically is 
that chiral symmetry can be realized in different ways.
To make this a bit more concrete,
let us go back to the continuum theory studied in the previous section.
As explained there,
the fermion propagator (4.12) along the domain wall
coincides with the left-handed components 
of the Green function of a modified Dirac operator $D$
[eq.~(4.13)].
Now the surprise is that $D$ does not
anticommute with $\dirac{5}$ and instead satisfies the equation
\equation{
  \dirac{5}D+D\dirac{5}={1\over M}D\dirac{5}D.
  \enum
}
This relation first appeared in 1982 in a 
paper of Ginsparg and Wilson [\ref{GinspargWilson}] 
on a completely different topic 
(the block-spin renormalization group 
in lattice QCD).
At the time it was considered to represent some sort 
of remnant chiral symmetry, but
there is in fact much more behind this remarkable identity.

\subsection 5.2 Significance of the Ginsparg--Wilson relation

In the following paragraphs we only use the Ginsparg--Wilson 
relation (5.1)
and the hermiticity property $D^{\dagger}=\dirac{5}D\dirac{5}$.
Although we are in the continuum theory here, 
the results that we shall obtain are of a very general nature
and can easily be carried over to the lattice (as we shall see
later in this section).

\vskip1ex
\noindent
(a)~{\sl Propagator}.
In terms of the Green function $S(x,y)$ of $D$, the 
Ginsparg--Wilson relation assumes the form
\equation{
  S(x,y)\dirac{5}+\dirac{5}S(x,y)=
  {1\over M}\kern1pt\dirac{5}\kern1pt\delta(x-y).
  \enum
}
The propagator is hence 
chirally invariant at all non-zero distances. 
In particular, the residue of the particle pole in momentum space
anticommutes with $\dirac{5}$ and we can say, therefore, that
the Ginsparg--Wilson relation implies chiral invariance on the mass shell.

\vskip1ex
\noindent
(b)~{\sl Off-shell chiral symmetry}. 
If we define the fermion action as usual,
\equation{
  \SF[\psibar,\psi]=\int\rmd^4x\,\psibar(x)D\psi(x),
  \enum
}
it is trivial to check that the infinitesimal transformation
\equation{
  \psi\to\psi+\epsilon\kern1pt\dirac{5}
  \left(1-D/M\right)\psi,
  \enum
  \next{2.5ex}
  \psibar\to\psibar+\epsilon\kern1pt\psibar\dirac{5},
  \enum
}
leaves the action invariant to first order in $\epsilon$ [\ref{ExactSymmetry}].
At momenta far below $M$, this transformation reduces to an ordinary
chiral rotation, and it can hence be regarded
as an unusual but exact realization of chiral symmetry. 
Note incidentally that the fermion and antifermion fields 
can be transformed independently of each other in euclidean space. 
Eventually they are both integrated over
in the functional integral, and the transformation (5.4),(5.5) is 
then just a substitution of integration variables.

\vskip1ex
\noindent
(c)~{\sl Weyl fermions}.
Now that we have an exact chiral symmetry, it is not difficult
to pass to chiral fermions. First note that the operator
$\dirachat=\dirac{5}(1-D/M)$ satisfies
\equation{
  (\dirachat)^{\dagger}=\dirachat,
  \qquad
  (\dirachat)^2=1,
  \qquad
  \dirac{5}D=-D\dirachat.
  \enum
}
The fermion action thus splits into left- and right-handed parts, if the 
chiral projectors for fermion and antifermion fields are defined through
[\ref{OverlapSplit},\ref{BoulderReview},\ref{AbelianChLGT}]
\equation{
  \Phat_{\pm}=\frac{1}{2}(1\pm\dirachat),
  \qquad
  P_{\pm}=\frac{1}{2}(1\pm\dirac{5}),
  \enum
}
respectively. It may seem strange not to use the 
same projectors for both fields, but in euclidean space this
is certainly permissible and probably also unavoidable 
in view of the transformation law (5.4),(5.5).

We can now eliminate the right-handed components by imposing the constraints
\equation{
  \Phat_{-}\psi=\psi,
  \qquad
  \psibar P_{+}=\psibar.
  \enum
}
The associated propagator is the inverse of 
the projected operator $P_{+}D\Phat_{-}$ 
in the subspace of left-handed fields and is thus given by
\equation{
  \Phat_{-}S(x,y)P_{+}=
  P_{-}S(x,y)P_{+}+{1\over2M}\kern1pt P_{+}\delta(x-y),
  \enum
}
where $S(x,y)$ denotes the Green function of $D$ as before.
Up to the contact term and an uninteresting normalization factor, 
this expression coincides with the propagator (4.12) of the 
fermion field in 4+1 dimensions along the domain wall. 
In other words, 
{\it at low energies, domain wall fermions reduce to
a chiral theory in four dimensions in which the chiral symmetry
is realized through the Ginsparg--Wilson relation}
[\ref{KikukawaNoguchi}].

\subsection 5.3 Lattice fermions with exact chiral symmetry

Domain wall fermions on a five-dimensional lattice 
can be studied in essentially
the same way as in the continuum theory. In particular, the domain wall
may again be replaced by a boundary with Dirichlet boundary conditions.
A further technical
simplification is achieved by taking the lattice spacing
in the extra dimension to zero while keeping the spacing $a$ 
in four dimensions fixed. The fermion propagator is then obtained
as before (subsect.~4.2). If we set 
\equation{
  D_5=D_{\rm w}+\dirac{5}\partial_s-M,
  \qquad
  M=1/a,
  \enum
}
for example, where $D_{\rm w}$ denotes the Wilson--Dirac operator (3.9),
the boundary value of the propagator is still given by
eq.~(4.12), with $S(x,y)$ the Green function
of the operator [\ref{OverlapDiracOperator}]
\equation{
  D={1\over a}\left\{1-(1-aD_{\rm w})
  \left[(1-aD_{\rm w})^{\dagger}(1-aD_{\rm w})\right]^{-1/2}
  \right\}.
  \enum
}
Up to lattice corrections of order $a$,
this operator coincides with $D_{\rm w}$, 
and it may hence 
be regarded as another possible discretization of the continuum Dirac operator.
Moreover, it satisfies the Ginsparg--Wilson relation,
\equation{
  \dirac{5}D+D\dirac{5}=aD\dirac{5}D,
  \enum
}
and also the hermiticity condition $D^{\dagger}=\dirac{5}D\dirac{5}$.
From our discussion in the previous subsection, we thus conclude that
lattice fermions in four dimensions with lattice Dirac operator $D$
preserve chiral symmetry in the form of the infinitesimal 
transformation (5.4),(5.5). In particular, as explained above, they can 
be split into left- and right-handed components in a consistent way.

Although this is not in conflict with the precise statement
of the Nielsen--Ninomiya theorem, 
some doubt may remain that we have indeed managed to 
avoid the doubling problem.
The following remarks should make it 
clear that there is really nothing to complain about here.

\vskip1ex
\noindent
(a)~{\sl Unphysical poles}.
From the definition (5.11) and eqs.~(3.15),(3.16) 
it is straightforward to deduce that $D$ is given by
\equation{
  D={1\over a}\left\{
  1-\bigl[1-\frac{1}{2}a^2\hat{p}^2-ia\dirac{\mu}\ring{p}_{\mu}\bigr]
  \bigl[\hbox{$1+\frac{1}{2}a^4
  \sum_{\mu<\nu}\hat{p}^2_{\mu}\hat{p}^2_{\nu}$}\bigr]^{-1/2}
  \right\}
  \enum
}
in momentum space. 
This expression is analytic in the whole  
range $|p_{\mu}|\leq\pi/a$ of lattice momenta.
It can also be shown to be invertible at all these momenta
except at $p=0$ where
\equation{
  D=i\dirac{\mu}p_{\mu}+\rmO(ap^2).
  \enum
}
In particular, the propagator $1/D$ has no unphysical poles.

\vskip1ex
\noindent
(b)~{\sl Locality}.
In position space the action of $D$ on an arbitrary fermion field $\psi(x)$  
is given by a kernel $D(x,y)$ through
\equation{
  D\psi(x)=a^4\sum_y D(x,y)\psi(y).
  \enum
}
The kernel is translation-invariant
and equal to the Fourier transform of the right-hand side
of eq.~(5.13). Starting from this representation, it is possible to derive
a bound of the form [\ref{Locality}]
\equation{
  \|D(x,y)\|\leq C\kern2pt\rme^{-\|x-y\|/\varrho},
  \enum
}
which shows that $D$ is a local operator with localization range $\varrho$.
From the point of view of 
the continuum limit, this is as good as strict locality,
since the range $\varrho$ is estimated to 
be no more than a few lattice spacings.

\vskip1ex
\noindent
(c)~{\sl Unitarity}.
As is generally true in a free field theory,
all the information about the underlying Hilbert space of physical states and 
the energy-momentum spectrum can be retrieved from the propagator
$S(x,y)$ (which represents $1/D$ in position space).
In particular, negative norm or complex energy states are excluded
if and only if the propagator admits a K\"all\'en--Lehmann representation
\equation{
  \left.S(x,y)\right|_{x_0>y_0}=
  \int_0^{\infty}\rmd E
  \int_{-\pi/a}^{\pi/a}{\rmd^3{\bf p}\over(2\pi)^3}\,
  \sigma(E,{\bf p})\kern1pt\rme^{-E(x_0-y_0)+i{\bf p(x-y)}}
  \enum
}
with non-negative spectral density, viz.
\equation{
   \bar{u}\sigma(E,{\bf p})u\geq0,
   \qquad
   \bar{u}=u^{\dagger}\dirac{0},
   \enum
}
for all Dirac spinors $u$. The expression (5.13) for 
the Dirac operator in momentum space looks rather complicated 
and it seems unlikely that $S(x,y)$ has such a representation, but
a lengthy exercise in complex contour integration shows this
to be the case.
Unitarity is hence respected for any value of the lattice spacing.

\subsection 5.4 Adding gauge fields

So far our discussion of domain wall fermions and the Ginsparg--Wilson
relation has been limited to free fermions.
Gauge fields may now easily be included, however, by 
replacing the free Dirac operator in four dimensions by the 
gauge-covariant operator. Most of the formulae
that we have obtained then remain valid without modification.
The fermion propagator in 4+1 dimensions, for example, is 
still given by eq.~(B.4), and 
the dimensional reduction thus works out as before.

In the presence of the gauge field, eq.~(5.11) defines
a gauge-covariant lattice Dirac operator that satisfies
the Ginsparg--Wilson relation (5.12).
Using this operator, we can build a version of lattice QCD in which
chiral symmetry is preserved. The infinitesimal chiral rotations
\equation{
  \psi\to\psi+\epsilon\kern1pt\lambda^a\dirachat\psi,
  \qquad
  \psibar\to\psibar+\epsilon\kern1pt\psibar\dirac{5}\lambda^a
  \enum
}
may include a flavour matrix $\lambda^a$ in this case,
and the lattice action
\equation{
  \SF[U,\psibar,\psi]=a^4\sum_x\psibar(x)D\psi(x)
  \enum
}
is then invariant under the full chiral flavour group.

This seems to contradict the fact that
the flavour-singlet axial symmetry 
in QCD with $\Nf$ massless quarks 
is known to be broken by the axial anomaly. 
However, symmetry transformations in quantum field theory
must not only preserve the action but also the integration 
measure in the functional integral.
In lattice QCD the fer\-mi\-on
and antifermion integration measures are given by
\equation{
  \rmD[\psi]=\prod_{x,\alpha}\rmd\psi_{\alpha}(x),
  \qquad
  \rmD[\psibar]=\prod_{x,\alpha}\rmd\smash{\hbox{$\displaystyle\psibar$}}
  _{\kern-0.4pt\alpha}(x),
  \enum
}
where $\alpha$ collectively denotes the Dirac, colour and flavour 
indices of the fields. To first order in $\epsilon$,
these measures transform according to 
\equation{
  \rmD[\psi]\to\bigl[1-\epsilon\kern1pt
  \Tr\{\lambda^a\dirachat\}\bigr]\kern1pt
  \rmD[\psi],
  \qquad
  \rmD[\psibar]\to\rmD[\psibar],
  \enum
}
under the chiral rotations (5.19). 

Since $\dirachat$ is trivial in 
flavour space, the jacobian in eq.~(5.22) is equal to unity~if
$\tr\{\lambda^a\}=0$.
The flavoured transformations are hence exact symmetries
of the theory.
This is not so in the flavour-singlet case, where
[\ref{IndexTheorem}--\ref{Adams}]
\equation{
  \Tr\left\{\dirachat\right\}=
  -a^4\sum_x\kern2pt\tr\left\{\dirac{5}aD(x,x)\right\}
  \noenum
  \next{2.0ex}
  \kern3.1em
  \mathrel{\mathop\sim_{a\to0}}
  -{\Nf\over32\pi^2}\int\rmd^4x\kern2pt 
  \epsilon_{\mu\nu\rho\sigma}
  F_{\mu\nu}^a(x)F_{\rho\sigma}^a(x).
  \enum
}
The axial anomaly thus arises
in the way suggested long ago by Fujikawa [\ref{FujikawaAnomaly}].
In particular, the symmetry structure in this formulation of lattice QCD 
is exactly as expected.

\section 6. Gauge-invariant lattice regularization of anomaly-free theories

Vector-like theories with any gauge group and 
fermion representation can be put on the lattice 
in the same manner,
using the appropriate gauge-covariant version of the 
lattice Dirac operator (5.11).
We can then pass to the associated chiral theory simply by imposing the 
constraints (5.8) on the fermion and antifermion fields.
As far as the field space, the lattice action and 
the classical field equations are concerned, 
the projection to the left-handed fields 
is completely consistent in this case, since
it derives from an underlying exact symmetry.

When we now try to quantize this theory through the functional integral,
the principal difficulty is that there is a 
non-trivial phase ambiguity in 
the fermion integration measure.
The problem is related to the gauge anomaly and
will occupy us throughout this section.
For simplicity we shall focus on the construction of the 
measure in perturbation theory and shall ignore 
all complications that have to do with non-perturbative effects
such as the global anomalies [\ref{Witten}--\ref{BaerCampos}].

Before plunging into the details, it is worth mentioning that
the fifth dimension has completely disappeared at this point.
We could in fact start from any gauge-covariant lattice Dirac operator that 
satisfies the Ginsparg--Wilson relation
and a few technical conditions 
(locality, absence of unphysical poles, etc.).
The operator (5.11) has all these properties, and also
the ``perfect" Dirac operator 
[\ref{PerfectDiracOperator},\ref{PerfectAction},\ref{PerfectActionReview}],
which derives from the block-spin renormalization group transformations
studied previously by Ginsparg and Wilson in their famous work
[\ref{GinspargWilson}].

\subsection 6.1 Functional integral

As usual the basic quantities to be considered in the quantized theory
are correlation functions 
$\langle \phi_1(x_1)\ldots\phi_n(x_n)\rangle$
of gauge-invariant local fields.
Formally they are given by the functional integral
\equation{
  \bigl\langle\phi_1(x_1)\ldots\phi_n(x_n)\bigr\rangle=
  \noenum
  \next{2.5ex}
  \quad{1\over{\cal Z}}
  \int\rmD[U]\int\rmD[\psi]_{\rm left}\rmD[\psibar]_{\rm left}\,
  \phi_1(x_1)\ldots\phi_n(x_n)\kern1pt
  \rme^{-\SG[U]-\SF[U,\psibar,\psi]},
  \enum
}
where $\rmD[U]$ denotes the standard integration measure for 
lattice gauge fields
and the normalization constant $\cal Z$ is defined through
the requirement that $\langle1\rangle=1$.

Since the measure $\rmD[U]$ and the gauge-field action $\SG[U]$ 
will not concern us here, there is no need to specify them explicitly.
It suffices to know that they are locally defined and that
they preserve the gauge and the lattice symmetries
[\ref{MunsterMontvayBook},\ref{RotheBook}].
The fermion action (5.20) is also local and invariant so that
a problem (if any) can only arise from 
the integration measures $\rmD[\psi]_{\rm left}$ and 
$\rmD[\psibar]_{\rm left}$. 

Let us now have a closer look at the fermion measure.
In terms of an orthonormal basis 
$v_j(x)$ of left-handed Dirac fields,
the fermion field may be written as
\equation{
  \psi(x)=\sum_jv_j(x)c_j.
  \enum
}
The coefficients $c_j$ in this expansion 
represent the independent degrees of 
freedom of the field, and a possible fermion measure is thus 
given by
\equation{
  \rmD[\psi]_{\rm left}=\prod_j\rmd c_j.
  \enum
}
There is in fact not much choice here, since
the coefficients $c_j$
are the generators of a Grassmann algebra (we are dealing with fermions).
The integration measure on any such algebra is unique up to 
a complex proportionality factor.
In particular, if we pass to a different orthonormal basis,
\equation{
  \tilde{v}_j(x)=\sum_lv_l(x)\left({\cal Q}^{-1}\right)_{lj},
  \qquad
  \tilde{c}_j=\sum_l{\cal Q}_{jl}c_l,
  \enum
}
the measure changes by the factor $\det{\cal Q}$,
which is a pure phase factor since the transformation matrix ${\cal Q}$ is 
unitary.

\topinsert
\vbox{
\vskip0.3cm
 
\centerline{\epsfxsize=4.5cm%
\epsfbox{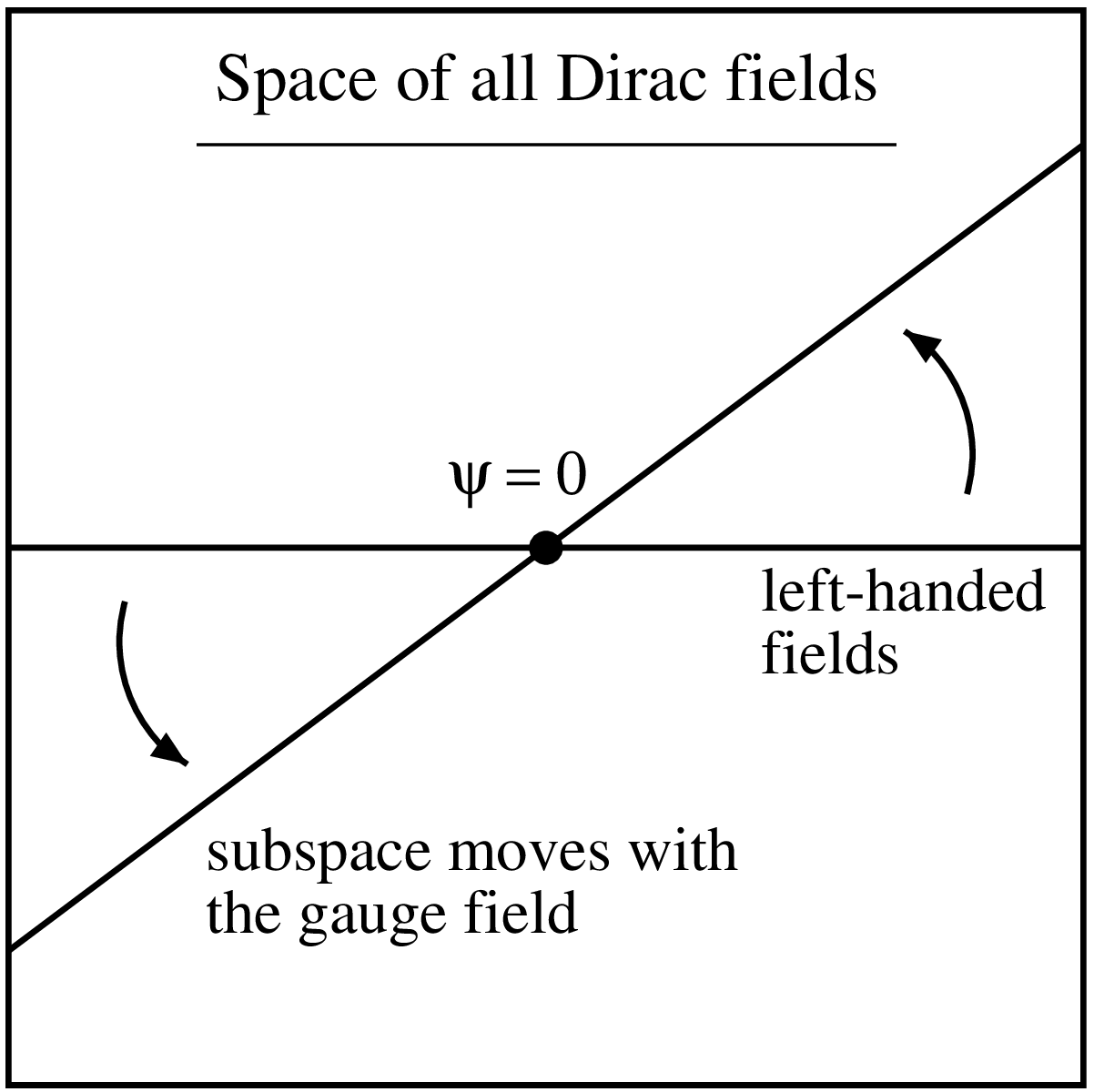}
}
\vskip0.3cm
\figurecaption{%
The projector $\Phat_{-}$ maps the space of all lattice Dirac fields to 
the subspace of left-handed fields. Since the projector 
involves the lattice Dirac operator $D$,
the~sub\-space changes when the gauge field is varied.
}
\vskip0.0cm
}
\endinsert

The antifermion measure $\rmD[\psibar]_{\rm left}$
is defined in the same way, 
using a basis $\bar{v}_k(x)$ of left-handed fields.
An important difference is that the basis can be taken 
to be independent of the gauge field, while this is not possible 
in the fermion case, because
the subspace of left-handed
fermion fields moves with the gauge field (see fig.~6). 
As a result the fermion measure and the partition function
\equation{
  \rme^{-\Seff[U]}=
  \int\rmD[\psi]_{\rm left}\rmD[\psibar]_{\rm left}\,
  \rme^{-\SF[U,\psibar,\psi]}
  \enum
}
have a gauge-field dependent phase ambiguity.
Evidently, the phase matters in the functional integral (6.1),
and the theory hence remains incompletely specified at the quantum level
until the ambiguity in the measure has been fixed.

Apart from this 
the structure of the theory has now been completely clarified.
In particular, since the integral over the fermion and antifermion fields 
is gaussian, the fermion correlation functions are given by
\equation{
  \int\rmD[\psi]_{\rm left}\rmD[\psibar]_{\rm left}\,
  \psi(x_1)\ldots\psi(x_n)\psibar(y_1)\ldots\psibar(y_n)
  \kern2pt\rme^{-S_{\rm F}[U,\psibar,\psi]}= 
  \noenum 
  \next{2.5ex}
  \kern4.0em
  \rme^{-\Seff[U]}\times
  \left\{\hbox{sum of Wick contractions}\right\}.
  \enum
}
The only non-zero two-point contraction is

\vbox{
  \vskip2.0ex
  \epsfxsize=2.2em\hskip1.3em%
  \epsfbox{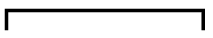}
  \vskip-6.68ex
  \equation{
    \psi(x)\psibar(y)=\hat{P}_{-}S(x,y)P_{+},
    \enum
  }
}

\vskip-1.0ex
\noindent
where $S(x,y)$ denotes the Green function of 
the lattice Dirac operator in the presence of the gauge field.
The fermion integral thus yields the expected result for the
chiral propagator (cf.~subsect.~5.2).

\subsection 6.2 Locality

The phase of the fermion measure 
should obviously be such that the locality properties and the symmetries of 
the theory are preserved. 
This requirement turns out to be very strong,
and it can be shown to 
determine the phase up to irrelevant local terms (i.e.~up to the 
usual discretization ambiguities)
[\ref{AbelianChLGT}--\ref{RegChGT}].
Of course, we may be unable to choose the phase in this way, 
but it is surely reasonable to try to get there.

We now need to say what locality precisely means 
in the present context. Directional derivatives in field space
are going to play an important r\^ole in this discussion, and 
we thus introduce these first. So let 
\equation{
  U_t(x,\mu)=\rme^{ta\eta_{\mu}\kern-1pt(x)}U(x,\mu),
  \qquad
  \eta_{\mu}(x)=\eta^a_{\mu}(x)T^a,
  \enum
}
be a smooth one-parameter family of lattice gauge fields (see fig.~7).
We can think of $U_t$ as a
curve in field space, and if $F[U]$ is any given 
functional, its rate of change along the curve at the point 
$U_0=U$ is
\equation{
  \delta_{\eta}F[U]=\left\{{\rmd\over\rmd t}F[U_t]\right\}_{t=0}.
  \enum
}
The differential operator $\delta_{\eta}$, which is defined through
this equation,
may be regarded as a derivative in the direction of the
tangential vector $\eta^a_{\mu}(x)$. 
It is analogous to the derivative 
\equation{
  \int\rmd^4x\,\eta^a_{\mu}(x){\delta\over\delta A^a_{\mu}(x)}
  \enum
}
in the continuum theory and in fact converges to this operator in 
the continuum limit.

\topinsert
\vbox{
\vskip0.3cm
 
\centerline{\epsfxsize=1.9cm%
\epsfbox{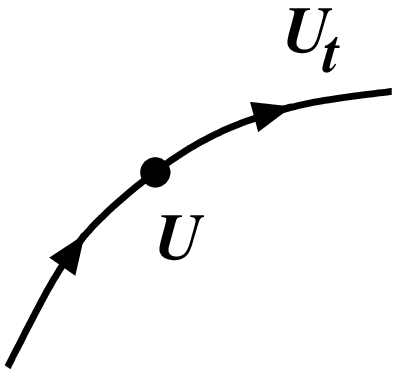}
}
\vskip0.2cm
\figurecaption{%
Equation (6.8) describes a curve in the space of all lattice gauge fields,
with curve parameter $t$ and tangential vector $\eta^a_{\mu}(x)$ at $t=0$.
}
\vskip0.0cm
}
\endinsert

The derivative of the effective action, 
$\delta_{\eta}\Seff[U]$, can be worked out by applying 
$\delta_{\eta}$ on both sides of eq.~(6.5).
Note that the fermion field in the integral is to be
replaced by the expansion (6.2). The fermion action then
depends on the gauge field not only through the Dirac operator but also
through the basis vectors $v_j(x)$. As a result the calculation 
yields two terms,
\equation{
  \delta_{\eta}\Seff=-\Tr\kern1pt\{
  \delta_{\eta}D\Phat_{-}D^{-1}P_{+}\}+i\L_{\eta},
  \enum
  \next{2.5ex}
  \L_{\eta}=i\sum_j\kern1pt(v_j,\delta_{\eta}v_j), 
  \enum
}
of which the second characterizes
the chosen measure and is hence referred to as the {\it measure term}. 
It depends linearly on the direction $\eta^a_{\mu}(x)$, i.e.~there exists
a current $j^a_{\mu}(x)$ such that
\equation{
  \L_{\eta}=
  a^4\sum_x\eta^a_{\mu}(x)j^a_{\mu}(x).
  \enum
}
At this point little is known about this current, except that 
it is a well-defined func\-tion of the gauge field
for any given choice of the fermion measure. 

If we temporarily consider the gauge field to be an external classical
field, eq.~(6.11) may be interpreted as the response of the 
fermion sector to a change of the field. 
The current $j^a_{\mu}(x)$ then appears on the 
right-hand side of the semiclassical field equations,
together with the other term, 
\equation{
  -\Tr\kern1pt\{
  \delta_{\eta}D\Phat_{-}D^{-1}P_{+}\}=
  a^4\sum_x\,\bigl\langle\psibar(x)\delta_{\eta}D\psi(x)\bigr\rangle_{\rm F},
  \enum
}
which can be written as a fermion expectation value of a local operator. 
A characteristic property of 
local theories is that the field equations are linear relations
between local operator insertions. 
We are thus led to require that {\it the current $j^a_{\mu}(x)$ should be
a local expression in the gauge field}. The measure term in eq.~(6.11)
then assumes the form of a local counterterm and 
the theory is only minimally affected by the 
gauge-field dependence of the fermion measure.

\subsection 6.3 Gauge invariance

The gauge transformation behaviour of the effective action can be studied
by computing its variation along the gauge directions in field space.
From the transformation law (3.10) 
it follows that these directions are given by
\equation{
  \eta_{\mu}(x)=-\nab{\mu}\omega(x),
  \qquad
  \omega(x)=\omega^a(x)T^a,
  \enum
}
where $\omega^a(x)$ is an arbitrary lattice field, transforming according to 
the adjoint representation of the gauge group, and $\nab{\mu}$ the 
appropriate covariant difference operator (cf.~subsect.~3.3).

The gauge variation of the effective action may now be worked out
by inserting eq.~(6.15) in eq.~(6.11). 
Using the gauge covariance of the Dirac operator, 
a few lines of algebra then lead to the result
\equation{
  \delta_{\eta}\Seff=ia^4\sum_x\kern1pt\omega^a(x)\bigl\{
  [\nabstar{\mu}j_{\mu}]^a(x)-\anomaly^a(x)\bigr\},
  \enum
  \next{2.0ex}
  \anomaly^a(x)=
  \frac{1}{2}\kern1pt ia\,\tr\{R(T^a)\dirac{5}D(x,x)\}.
  \enum
}
Since the kernel $D(x,y)$ of the Dirac operator 
depends locally on the gauge field [\ref{Locality}], 
it is obvious that $\anomaly^a(x)$ is a gauge-covariant local composite field.
In the continuum limit we have
[\ref{KikukawaYamada}--\ref{Adams},\ref{NonAbelianChLGT}]
\equation{
  \anomaly^a(x)\mathrel{\mathop\sim_{a\to0}}
  -{1\over128\pi^2}\kern1pt d^{abc}_R\epsilon_{\mu\nu\rho\sigma}
  F^b_{\mu\nu}(x)F^c_{\rho\sigma}(x)+\rmO(a),
  \enum
}
and $\anomaly^a(x)$ may hence be regarded as a lattice representation 
of the so-called covariant gauge anomaly
[\ref{Luis},\ref{Bertlmann}]. 

To ensure the gauge invariance of the effective action
(and consequently of the full theory),
the right-hand side of eq.~(6.16) must vanish for
all gauge variations. The fermion measure should hence be such that
the associated current $j^a_{\mu}(x)$ satisfies
\equation{
  [\nabstar{\mu}j_{\mu}]^a(x)=\anomaly^a(x).   
  \enum
}
Since the anomaly is a local field,
this is consistent with the requirement of locality, although 
there is obviously no guarantee that both conditions can be met.

\subsection 6.4 Integrability condition

Apart from locality and gauge invariance,
there is a further constraint on the current $j^a_{\mu}(x)$ 
that derives from the fact that eq.~(6.11) must
be integrable. In other words, if the equation is integrated 
along any smooth path $U_t$ in field space, 
the integral should only depend on the initial and final field configurations,
but not on the shape of the curve. This will be the case if and only if the 
curl in field space of the right-hand side of eq.~(6.11) vanishes.

Using the identities (5.6), this leads to the equation
\equation{
  \delta_{\eta}\L_{\zeta}-\delta_{\zeta}\L_{\eta}+a\L_{[\eta,\zeta]}
  =i\kern2pt\Tr\{\hat{P}_{-}
  [\delta_{\eta}\hat{P}_{-},\delta_{\zeta}\hat{P}_{-}]\},
  \enum
}
where the commutator of the two vector fields
$\eta^a_{\mu}(x)$ and $\zeta^a_{\mu}(x)$ is to be taken point-wise.
Note that the right-hand side of eq.~(6.20) is 
a known local function of these fields and the gauge field,
while the other side involves the measure term only.
The integrability condition thus has the form of an inhomogeneous
partial differential
equation in field space for the current $j^a_{\mu}(x)$.

So far the fermion measure has been taken as the primary quantity
from which the associated current is obtained. 
We can now invert this relationship since
{\it any given current
that satisfies the integrability condition arises from
an underlying fermion measure} [\ref{NonAbelianChLGT}].
In other words, to complete the construction of the theory, 
it suffices to find a current $j^a_{\mu}(x)$ that (a) is a local expression
in the gauge field, (b) fulfils the requirement of gauge invariance, 
eq.~(6.19), and (c) solves the integrability condition (6.20).
This does not look like an easy task, but the problem is 
in fact much more accessible than it seems
to be, at least in perturbation theory.

\subsection 6.5 Construction of the fermion measure in perturbation theory

The perturbation expansion of the functional integral (6.1) around the 
vacuum configuration $U(x,\mu)=1$ is obtained essentially as in lattice QCD.
An explicit specification of the fermion measure is not required
if we first integrate over the fermion fields 
using eqs.~(6.6),(6.7).
The integral then assumes the form
\equation{
  \bigl\langle\phi_1(x_1)\ldots\phi_n(x_n)\bigr\rangle=
  {1\over{\cal Z}}
  \int\rmD[U]\,
  \bigl\langle
  \phi_1(x_1)\ldots\phi_n(x_n)
  \bigr\rangle_{\rm F}
  \kern1pt\rme^{-\SG[U]-\Seff[U]}.
  \enum
}
As usual the gauge needs to be fixed and 
the lattice field $U(x,\mu)$ is para\-metrized through 
a gauge potential $A^a_{\mu}(x)$ according to eq.~(3.13).
After rescaling 
\equation{
  A^a_{\mu}(x)\to g_0A^a_{\mu}(x)
  \enum
}
(where $g_0$ denotes the bare gauge coupling),
the perturbation series is obtained by 
expanding the integrand in eq.~(6.21) in powers of $g_0$.

The only really new element here is the 
measure term, which [via eq.~(6.11)] con\-tri\-butes
to the expansion of the effective action.
In perturbation theory
the associated current is given by
\equation{
  j^a_{\mu}(x)=\sum_{n=0}^{\infty}
  {g_0^n\over n!}\kern1pt a^{4n}\kern-3pt\sum_{x_1,\ldots,x_n}\kern-2pt
  L^{(n)}(x,x_1,\ldots,x_n)^{aa_1\ldots a_n}_{\mu\mu_1\ldots\mu_n}
  A^{a_1}_{\mu_1}(x_1)\ldots A^{a_n}_{\mu_n}(x_n),
  \enum
}
and we are now left with the problem of determining the coefficients 
$L^{(n)}$ so that the conditions listed above are satisfied.
The expansion of the 
effective action can then be worked out straightforwardly 
by repeated differentiation of eq.~(6.11)
with respect to the gauge potential.

It may be useful at this point to briefly state what the conditions
(a)--(c) mean in terms of the coefficients $L^{(n)}$. 
To fulfil the requirement of locality, 
the coefficients must be translation-invariant and exponentially
decreasing if any of the distances $\|x_k-x\|$ becomes
larger than a few lattice spacings.
The other conditions both turn into towers of 
inhomogeneous linear equations for the coefficients
once the anomalous conservation law (6.19) and 
the integrability condition (6.20) are ex\-pan\-ded 
in powers of the gauge potential.
Although this rapidly becomes very tedious,
it is possible to work out the right-hand sides of these
equations explicitly by expressing them through
the kernel $D(x,y)$ of the Dirac operator and expanding the latter
following 
refs.~[\ref{KikukawaYamada},\ref{RegChGT},%
\ref{AlexandrouEtAlI}--\ref{Capitani}].

It is surely far from obvious that all these equations can be solved,
but it turns out that a solution can be constructed through a 
purely algebraic recursive procedure [\ref{RegChGT}].  
As a result we have the following

\proclaim Theorem.
If the fermion representation $R$ is anomaly-free, there exist lattice 
functions $L^{(n)}$ such that 

\vskip-0.5ex\noindent
{\sl(1) $j^a_{\mu}(x)$ satisfies conditions (a)--(c)
to all orders of $g_0$,}

\vskip1.3ex\noindent
{\sl(2) $L^{(n)}=0$ for $n\leq3$,}

\vskip1.3ex\noindent
{\sl(3) $\Im\Seff$ transforms like a pseudoscalar under 
the lattice symmetries.}

\vskip1.5ex
\noindent
While there is more than one solution with these properties,
the difference between any two of them amounts to a change 
of the effective action by a term of the form
\equation{
  \Delta\Seff[U]=
  a^4\sum_x\Omega(x), 
  \enum
}
where $\Omega(x)$ is a 
gauge-invariant, pseudoscalar local field of dimension 
greater than or equal to 5. 
For dimensional reasons,
$\Omega(x)$ must be proportional to a positive power of the lattice spacing,
and up to finite renormalizations
such differences are hence not expected to have any influence on the theory 
in the continuum limit.

At this point the lattice theory has been completely specified, to all
orders of the gauge coupling. The fact that we have been able 
to meet all conditions on the fermion measure shows that
{\it anomaly-free chiral gauge theories can be consistently
regularized without breaking the gauge symmetry}. 
While the construction of this re\-gu\-la\-ri\-zation is obviously non-trivial, 
it should be emphasized 
that no computation of Feynman diagrams was required.
The equations that had to be solved are explicitly given 
and involve local expressions only.

\subsection 6.6 Anomaly cancellation \& cohomology

The demonstration of the exact cancellation of the gauge anomaly
on the lattice is perhaps the most difficult step
in the proof of the theorem quoted above. 
We shall not attempt to describe this in any detail, 
but a quick look at the problem 
and its solution in the abelian case may be illuminating.

So let us consider a U(1) theory with fermion representation (2.16).
Since there is only one group generator,  
the anomaly (6.17) has no index and is given by
\equation{
  \anomaly(x)=\frac{1}{2}\kern1pt ia\kern1pt\tr\{R\dirac{5}D(x,x)\},
  \qquad
  R=i\times\hbox{diag}(\rme_1,\ldots,\rme_N).
  \enum
}
In the present context we may assume that 
$\anomaly(x)$ is expanded in powers of the gauge potential 
$A_{\mu}(x)$ similarly to the current $j_{\mu}(x)$
[eq.~(6.23)]. 
From eq.~(6.25) we then infer that 
the associated coefficients are local and that
the series is invariant under arbitrary gauge transformations
\equation{
  A_{\mu}(x)\to A_{\mu}(x)+g_0^{-1}\drv{\mu}\omega(x)
  \enum
}
(where $\drv{\mu}$ denotes the forward difference operator).

A less obvious property of the anomaly is that
\equation{
  a^4\sum_x\delta\anomaly(x)=0
  \enum
}
for any local variation $\delta A_{\mu}(x)$ of the gauge potential.
To prove this we note that the left-hand side of eq.~(6.27)
is proportional to $\Tr\{R\delta\dirachat\}$. 
We then insert $(\dirachat)^2=1$
and cycle one of the factors $\dirachat$
around the trace using
\equation{
  \{\dirachat,\delta\dirachat\}=0,
  \qquad [\kern1pt\dirachat,R\kern1pt]=0.
  \enum
}
As a result the trace is reproduced
with the opposite sign and it must hence be equal to zero.

Equation (6.27) says that 
the abelian anomaly is a {\it local topological field},
i.e.~it has all the characteristic 
properties of a topological density.
Modulo divergence terms 
(which are topologically uninteresting)
there are usually not many fields of this type,
and the form of the anomaly is thus strongly constrained.
The classification of topological fields
is a particular case of a local cohomology problem,
a subject that has received a lot of attention 
in continuum field theory. 
In particular, using the descent equations [\ref{StoraI}--\ref{BaulieuII}],
a general theorem has been established
which states that, in pure gauge theories 
with any gauge group, the Chern monomials are the only
non-trivial topological fields
[\ref{BrandtEtAl},\ref{DuboisVioletteEtAl}].

On the lattice, a similar classification theorem holds
for U(1) theories [\ref{AbelianCohomologyI},\ref{AbelianCohomologyII}].
Any topological field $q(x)$
can be shown to be of the form
\equation{
  q(x)=c(x)+\drvstar{\mu}k_{\mu}(x)
  \enum
}
in this case,
where $k_{\mu}(x)$ is a {\it gauge-invariant local}\/ current and
\equation{
  c(x)=\alpha+\beta_{\mu\nu}F_{\mu\nu}(x)+
  \gamma\epsilon_{\mu\nu\rho\sigma}
  F_{\mu\nu}(x)F_{\rho\sigma}(x+a\hat{\mu}+a\hat{\nu}),
  \enum
  \next{2.5ex}
  F_{\mu\nu}(x)=\drv{\mu}A_{\nu}(x)-\drv{\nu}A_{\mu}(x),
  \enum
}
the general Chern polynomial on the lattice
\footnote\dag{\footnotefont%
In the last term in eq.~(6.30)
the coordinate shift is required to ensure 
that $c(x)$ is topological. 
Its presence can be traced back to the fact 
that the difference operators $\drv{\mu}$ and $\drvstar{\mu}$
obey a modified Leibniz rule [\ref{AbelianCohomologyII}].}.
The proof of the theorem is constructive
in the sense that the coefficients of the 
current $k_{\mu}(x)$ are obtained algebraically in terms of
the coefficients of $q(x)$.

In the case of the anomaly,
the constants $\alpha$ and $\beta_{\mu\nu}$
must vanish, because $\anomaly(x)$ transforms like a 
pseudoscalar field under the lattice symmetries. 
Concerning the constant $\gamma$ we note that
the anomaly is a sum of terms, one for each fermion flavour.
Since the field tensor scales with the charge and 
since there is another power of the charge coming from the 
representation matrix $R$ in eq.~(6.25), we have
\equation{ 
  \gamma\propto\sum_{\alpha=1}^N\rme_{\alpha}^3.
  \enum
}
The topologically non-trivial part of the anomaly
thus cancels if the fermion representation is anomaly-free,
and the condition for exact gauge invariance, eq.~(6.19),
then reduces to 
\equation{
  \drvstar{\mu}j_{\mu}(x)=\drvstar{\mu}k_{\mu}(x).
  \enum
}
Although more work is required to actually 
show this [\ref{AbelianChLGT},\ref{RegChGT}],
it should now be quite plausible that a current $j_{\mu}(x)$
can be found that satisfies this equation and all 
the other conditions. At least the topological
obstruction represented by the anomaly has completely
disappeared at this stage.

\subsection 6.7 Epilogue

In order to bring out the basic ideas as clearly as possible,
many technical issues have been left aside in this section
and the presentation has often been rather abbreviated.
A much more detailed description of the formulation of chiral lattice
gauge theories in perturbation theory, to the extent that explicit
computations could start from there,
can be found in ref.~[\ref{RegChGT}]. 

At the non-perturbative level
a general construction of the fermion measure is still missing.
The principal difficulties are that there can be global obstructions
and that the local constraints are less accessible
if the equations may not be expanded in powers of the gauge potential
[\ref{NonAbelianChLGT}].
Currently all these mathematical problems have been completely solved
only for abelian gauge groups [\ref{AbelianChLGT}].

\topinsert
\vbox{
\vskip0.3cm
 
\centerline{\epsfxsize=6.6cm%
\epsfbox{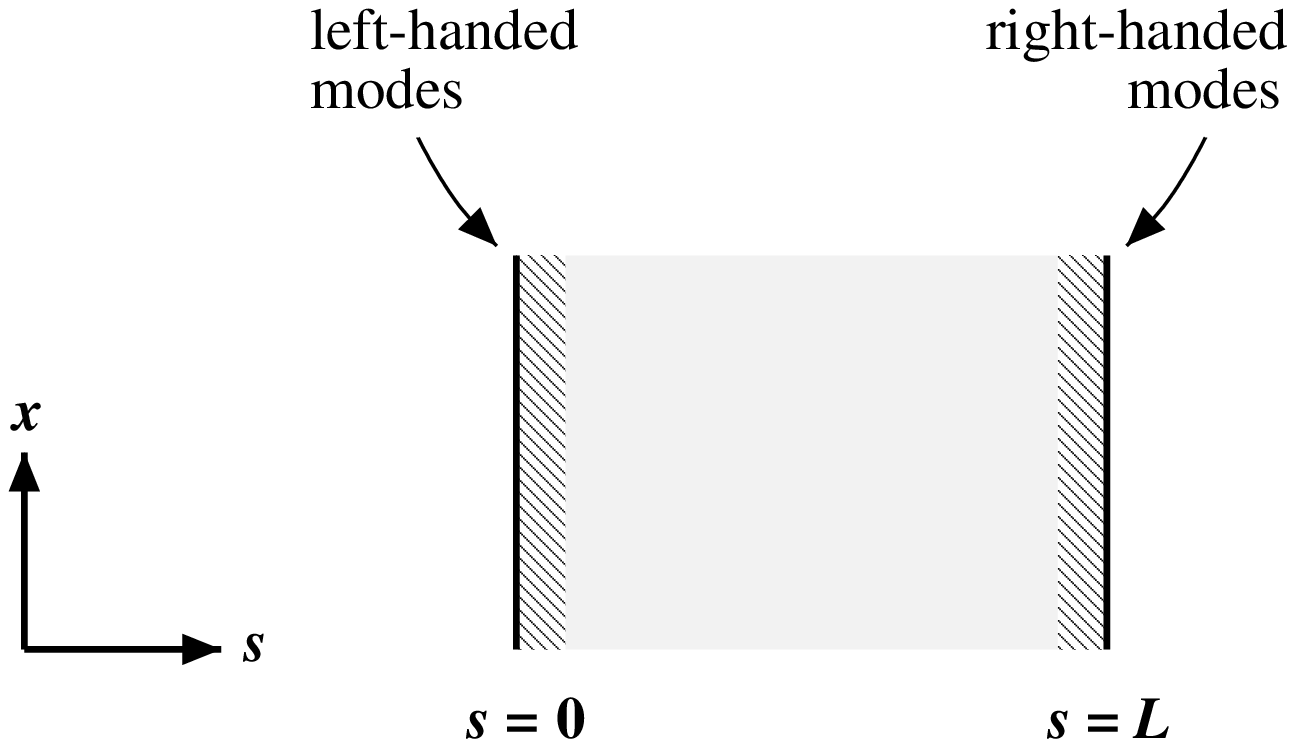}
}
\vskip0.2cm
\figurecaption{%
At low energies,
massive fermions in a five-dimensional volume with two boundaries
reduce to left- and right-handed fermions in four dimensions
that couple to the gauge field
in the vicinity of the boundaries.
}
\vskip1.0ex
}
\endinsert

In these lectures we made a long excursion to five dimensions,
which led us to the Ginsparg--Wilson relation and a new
realization of chiral symmetry. 
This proved to be the key to the 
construction of chiral lattice gauge
theories in four dimensions.
The question may now be asked whether chiral gauge theories
can also be obtained directly from a field theory in five 
(or more) dimensions.
We might consider a five-dimensional volume
with two boundaries, for example, and couple 
an $s$-dependent gauge field $U(x,s,\mu)$ to a massive Dirac field
(see fig.~8). 
There are massless fermion modes with both chiralities
in this case, which interact with the
gauge field in the vicinity of the boundaries.
So if we impose the boundary conditions
\equation{
  \left.U(x,s,\mu)\right|_{s=0}=U(x,\mu),
  \qquad
  \left.U(x,s,\mu)\right|_{s=L}=1,
  \enum
}
the right-handed modes are expected to decouple for large 
wall separations $L$, 
while the left-handed
modes couple to the physical gauge field $U(x,\mu)$ only. 

This is in fact what happens
if the gauge field is treated as an external
classical field and if the fermion representation is anomaly-free
[\ref{AlvarezEtAl}--\ref{KikukawaAoyama}].
For several reasons it remains unclear, however,
whether a positive answer 
to the question posed above can be given along these lines.
The main problem at present is that the gauge field in 
the bulk of the five-dimensional volume needs to be decoupled
from the physical field at the boundary, and there seems to be
no natural way to achieve this.

\ninepoint

\beginbibliography

% Reviews on anomalies, descent equations etc.

\bibitem{Luis}
L. Alvarez--Gaum\'e,
An introduction to anomalies, 
in: Fundamental problems of gauge field theory (Erice 1985),
eds. G. Velo, A. S. Wightman (Plenum Press, New York, 1986)

\bibitem{Bertlmann}
R. A. Bertlmann, Anomalies in quantum field theory (Oxford University
Press, Oxford, 1996)

% Algebraic renormalization

\bibitem{BRS}
C. Becchi, A. Rouet, R. Stora, 
Commun. Math. Phys. 42 (1975) 127;
Ann. Phys. (NY) 98 (1976) 287

\bibitem{Becchi}
C. Becchi,
Lectures on the renormalization of gauge theories,
in: Relativity, groups and topology 
(Les Houches 1983), 
eds. B. S. DeWitt, R. Stora
(North--Holland, Amsterdam, 1984)

\bibitem{PiguetSorella}
O. Piguet, S. P. Sorella,
Algebraic renormalization: 
perturbative renormalization, symmetries and anomalies,
Lecture Notes in Physics m28
(Springer--Verlag, Berlin, 1995)

% Global anomalies

\bibitem{Witten}
E. Witten,
Phys. Lett. B117 (1982) 324;
Nucl. Phys. B223 (1983) 422

\bibitem{ElitzurNair}
S. Elitzur, V. P. Nair,
Nucl. Phys. B243 (1984) 205

\bibitem{BaerCampos}
O. B\"ar, I. Campos,
Nucl. Phys. B (Proc. Suppl.) 83--84 (2000) 594;
Nucl. Phys. B581 (2000) 499

% `Rome approach' to chiral gauge theories

\bibitem{Rome}
A. Borrelli, L. Maiani, G. C. Rossi, R. Sisto, M. Testa,
Phys. Lett. B221 (1989) 360;
Nucl. Phys. B333 (1990) 335

\bibitem{RomeReviewI}
M. Testa, 
The Rome approach to chirality,
in: Recent developments in non-perturbative quantum field theory
(Seoul 1997), eds. Y. M. Cho, M. Virasoro 
(World Scientific, Singapore, 1998)

\bibitem{AlonsoEtAlI}
J. L. Alonso, J. L. Cortes, E. Rivas,
Int. J. Mod. Phys. A5 (1990) 2839

\bibitem{AlonsoEtAlII}
J. L. Alonso, P. Boucaud, F. Lesmes, A. J. van der Sijs,
Nucl. Phys. B457 (1995) 175

\bibitem{BockEtAlI}
W. Bock, M. F. L. Golterman, Y. Shamir,
Nucl. Phys. (Proc. Suppl.) 63 (1998) 147 and 581;
Phys. Rev. Lett. 80 (1998) 3444;
Phys. Rev. D58 (1998) 034501

\bibitem{BockEtAlII}
W. Bock, M. F. L. Golterman, K. C. Leung, Y. Shamir,
Nucl. Phys. B (Proc. Suppl.) 83--84 (2000) 603;
Chin. J. Phys. 38 (2000) 647

\bibitem{RomeReviewII}
M. Testa, 
Chin. J. Phys. 38 (2000) 563

% Application of the algebraic renormalization method

\bibitem{GrassiHurth}
P. A. Grassi, T. Hurth,
hep-ph/0101183

% Gauge-invariant operators satisfying the GW relation

\bibitem{PerfectDiracOperator}
P. Hasenfratz,
Nucl. Phys. B (Proc. Suppl.) 63A-C (1998) 53;
Nucl. Phys. B525 (1998) 401 

\bibitem{OverlapDiracOperator}
H. Neuberger,
Phys. Lett. B417 (1998) 141;
{\it ibid.}\/ B427 (1998) 353

% Exact index theorem

\bibitem{IndexTheorem}
P. Hasenfratz, V. Laliena, F. Niedermayer,
Phys. Lett. B427 (1998) 125

% Exact chiral symmetry

\bibitem{ExactSymmetry}
M. L\"uscher,
Phys. Lett. B428 (1998) 342

% Locality of the Neuberger's lattice Dirac operator

\bibitem{Locality}
P. Hern\'andez, K. Jansen, M. L\"uscher,
Nucl. Phys. B552 (1999) 363

% Anomaly coefficient computed in perturbation theory

\bibitem{KikukawaYamada}
Y. Kikukawa, A. Yamada,
Phys. Lett. B448 (1999) 265 

\bibitem{Chiu}
T.--W. Chiu, 
Phys. Lett. B445 (1999) 371

\bibitem{ReiszRothe}
T. Reisz, H. J. Rothe,
Phys. Lett. B455 (1999) 246;
Nucl. Phys. B575 (2000) 255

% Classical continuum limit of the covariant anomaly

\bibitem{Fujikawa}
K. Fujikawa,
Nucl. Phys. B546 (1999) 480

\bibitem{Suzuki}
H. Suzuki,
Prog. Theor. Phys. 102 (1999) 141

\bibitem{Adams}
D. H. Adams, hep-lat/9812003;
Nucl. Phys. B589 (2000) 633

% Splitting of Dirac fields into chiral components

\bibitem{OverlapSplit}
R. Narayanan,
Phys. Rev. D58 (1998) 97501

\bibitem{BoulderReview}
F. Niedermayer,
Nucl. Phys. B (Proc. Suppl.) 73 (1999) 105

% Classification of topological fields on the lattice (Abelian case)

\bibitem{AbelianCohomologyI}
M. L\"uscher,
Nucl. Phys. B538 (1999) 515

\bibitem{AbelianCohomologyII}
T. Fujiwara, H. Suzuki, K. Wu,
Phys. Lett. B463 (1999) 63;
Nucl. Phys. B569 (2000) 643

% Anomaly cancellation for gauge group SU(2)xU(1)

\bibitem{KikukawaNakayama}
Y. Kikukawa, Y. Nakayama,
hep-lat/0005015

% Chiral gauge theory on the lattice with exact gauge invariance

\bibitem{AbelianChLGT}
M. L\"uscher,
Nucl. Phys. B549 (1999) 295

\bibitem{SuzukiAbelianChLGT}
H. Suzuki,
Prog. Theor. Phys. 101 (1999) 1147

\bibitem{NonAbelianChLGT}
M. L\"uscher,
Nucl. Phys. B568 (2000) 162

\bibitem{SuzukiBRS}
H. Suzuki,
Nucl. Phys. B585 (2000) 471
[E: H. Igarashi, K. Okuyama, H. Suzuki, hep-lat/0012018]

\bibitem{RegChGT}
M. L\"uscher,
J. High Energy Phys. 06 (2000) 028

\bibitem{SuzukiRealRep}
H. Suzuki,
J. High Energy Phys. 10 (2000) 039

% Original papers on descent equations

\bibitem{StoraI}
R. Stora,
Continuum gauge theories,
in: New developments in quantum field theory and statistical mechanics
(Carg\`ese 1976),
eds. M. L\'evy, P. Mitter (Plenum Press, New York, 1977)

\bibitem{StoraII}
R. Stora,
Algebraic structure and topological origin of anomalies,
in: Progress in gauge field theory 
(Carg\`ese 1983),
eds. G. 't Hooft et al. (Plenum Press, New York, 1984)

\bibitem{Zumino}
B. Zumino,
Chiral anomalies and differential geometry,
in: Relativity, groups and topology 
(Les Houches 1983), 
eds. B. S. DeWitt, R. Stora
(North--Holland, Amsterdam, 1984)

\bibitem{BaulieuI}
L. Baulieu,
Algebraic construction of gauge invariant theories,
in: Particles and Fields (Carg\`ese 1983),
eds. M. L\'evy et al. (Plenum, New York, 1985)

\bibitem{BaulieuII}
L. Baulieu,
Nucl. Phys. B241 (1984) 557;
Phys. Rep. 129 (1985) 1

% Chiral fermions from 4+1 dimensions

\bibitem{RubakovShaposhnikov}
V. Rubakov, M. Shaposhnikov,
Phys. Lett. B125 (1983) 136

\bibitem{CallanHarvey}
C. G. Callan, J. A. Harvey,
Nucl. Phys. B250 (1985) 427

\bibitem{Kaplan}
D. B. Kaplan,
Phys. Lett. B288 (1992) 342;
Nucl. Phys. B (Proc. Suppl.) 30 (1993) 597

\bibitem{Shamir}
Y. Shamir, 
Nucl. Phys. B406 (1993) 90

\bibitem{FurmanShamir}
V. Furman, Y. Shamir,
Nucl. Phys. B439 (1995) 54

\bibitem{KikukawaNoguchi}
Y. Kikukawa, T. Noguchi,
hep-lat/9902022

% 4+1 dimensional representation of the chiral fermion determinant

\bibitem{AlvarezEtAl}
L. Alvarez--Gaum\'e, S. Della Pietra, V. Della Pietra,
Phys. Lett. B166 (1986) 177;
Commun. Math. Phys. 109 (1987) 691

\bibitem{NielsBohr}
L. Alvarez--Gaum\'e, S. Della Pietra,
The effective action for chiral fermions,
in: Recent developments in quantum field theory
(Niels Bohr Centennial Conference, Copenhagen, 1985),
eds. J. Ambj{\o}rn et al. (North--Holland, Amsterdam, 1985)

\bibitem{BallOsborn}
R. D. Ball, H. Osborn,
Phys. Lett. B165 (1985) 410;
Nucl. Phys. B263 (1986) 245

\bibitem{Ball}
R. D. Ball,
Phys. Lett. B171 (1986) 435;
Phys. Rep. 182 (1989) 1

\bibitem{KaplanSchmaltz}
D. B. Kaplan, M. Schmaltz,
Phys. Lett. B368 (1996) 44

\bibitem{SuzukiAction}
H. Suzuki,
Prog. Theor. Phys. 101 (1999) 1147

\bibitem{KikukawaAoyama}
T. Aoyama, Y. Kikukawa, 
hep-lat/9905003

% Ginsparg-Wilson relation

\bibitem{GinspargWilson}
P. H. Ginsparg, K. G. Wilson,
Phys. Rev. D25 (1982) 2649

% Reviews of chiral LGT

\bibitem{ShamirReview}
Y. Shamir, 
Nucl. Phys. B (Proc. Suppl.) 47 (1996) 212

\bibitem{LuscherReview}
M. L\"uscher,
Nucl. Phys. B (Proc. Suppl.) 83--84 (2000) 34

\bibitem{NeubergerReview}
H. Neuberger,
Nucl. Phys. B (Proc. Suppl.) 83--84 (2000) 67

\bibitem{GoltermanReview}
M. F. L. Golterman,
hep-lat/0011027

% Chiral determinant in the continuum (proper time regularization)

\bibitem{Leutwyler}
H. Leutwyler,
Phys. Lett. B152 (1985) 78

% SU(5) unification of the strong and electroweak interactions

\bibitem{SUfive}
H. Georgi, S. L. Glashow,
Phys. Rev. Lett. 32 (1974) 438

% Compact Lie groups (classification of invariant tensors)

\bibitem{Zelobenko}
D. P. \v{Z}elobenko,
Compact Lie groups and their representations
(American Mathema\-ti\-cal Society, Providence, 1973)

% Dimensional regularization

\bibitem{DimRegI}
G. 't Hooft, M. Veltman,
Nucl. Phys. B44 (1972) 189

% Regularization with higher-derivative terms

\bibitem{SlavnovI}
A. A. Slavnov,
Nucl. Phys. B31 (1971) 301

\bibitem{LeeZinn}
B. W. Lee, J. Zinn--Justin,
Phys. Rev. D5 (1972) 3121

\bibitem{Slavnov}
A. A. Slavnov,
Theor. Math. Phys. 33 (1977) 977

\bibitem{FaddeevSlavnov}
L. D. Faddeev, A. A. Slavnov,
Gauge fields: introduction to quantum theory, 2nd ed.
(Benjamin--Cummings, London, 1991) 

\bibitem{MartinEtAlI}
C. P. Martin, F. Ruiz Ruiz,
Nucl. Phys. B436 (1995) 545

\bibitem{MartinEtAlII}
J. H. Le\'on, C. P. Martin, F. Ruiz Ruiz,
Phys. Lett. B355 (1995) 531

\bibitem{AsoreyFalceto}
M. Asorey, F. Falceto,
Phys. Rev. D54 (1996) 5290

\bibitem{BakeyevSlavnov}
T. D. Bakeyev, A. A. Slavnov,
Mod. Phys. Lett. A11 (1996) 1539

\bibitem{DeminovSlavnov}
M. M. Deminov, A. A. Slavnov,
hep-th/0012138

% Standard formulation of lattice QCD

\bibitem{Wilson}
K. G. Wilson, Phys. Rev. D10 (1974) 2445

% Textbooks on LGT 

\bibitem{MunsterMontvayBook}
I. Montvay, G. M\"unster,
Quantum fields on a lattice
(Cambridge University Press, Cambridge, 1994)

\bibitem{RotheBook}
H. J. Rothe,
Lattice gauge theories: an introduction, 2nd ed.
(World Scientific, Singapore, 1997)

% Gauge fixing on the lattice, renormalization & continuum limit

\bibitem{LesHouches}
M. L\"uscher,
Selected topics in lattice field theory,
in: Fields, strings and 
critical phenomena (Les Houches 1988), eds. E. Br\'ezin, J. Zinn--Justin 
(North--Holland, Amsterdam, 1989)

\bibitem{Reisz}
T. Reisz,
Commun. Math. Phys. 116 (1988) 81 and 573; 
{\it ibid.}\/ 117 (1988) 79 and 639;
Nucl. Phys. B318 (1989) 417

\bibitem{BFM}
M. L\"uscher, P. Weisz,
Nucl. Phys. B452 (1995) 213

% Nielsen-Ninomiya no-go theorem

\bibitem{NN}
H. B. Nielsen, M. Ninomiya,
Phys. Lett. B105 (1981) 219;
Nucl. Phys. B185 (1981) 20 [E: {\it ibid.}\/ B195 (1982) 541];
{\it ibid.}\/ B193 (1981) 173

\bibitem{Friedan}
D. Friedan,
Commun. Math. Phys. 85 (1982) 481

% Fujikawa's method

\bibitem{FujikawaAnomaly}
K. Fujikawa,
Phys. Rev. Lett. 42 (1979) 1195;
Phys. Rev. D21 (1980) 2848 [E: {\it ibid.} D22 (1980) 1499]

% Perfect action approach

\bibitem{PerfectAction}
P. Hasenfratz, F. Niedermayer,
Nucl. Phys. B414 (1994) 785

\bibitem{PerfectActionReview}
P. Hasenfratz,
The theoretical background and properties of perfect actions,
in: Non-perturbative quantum field physics (Pe\~niscola 1997),
eds. M. Asorey, A. Dobado (World Scientific, Singapore, 1998)

% Expansion of N's operator in perturbation theory

\bibitem{AlexandrouEtAlI}
C. Alexandrou, H. Panagopoulos, E. Vicari,
Nucl. Phys. B571 (2000) 257

\bibitem{IshibashiEtAl}
M. Ishibashi, Y. Kikukawa, T. Noguchi, A. Yamada,
Nucl. Phys. B576 (2000) 501

\bibitem{AlexandrouEtAlII}
C. Alexandrou, E. Follana, H. Panagopoulos, E. Vicari,
Nucl. Phys. B580 (2000) 394

\bibitem{Capitani}
S. Capitani, 
Nucl. Phys. B592 (2000) 183

% Classification of topological fields in the continuum theory

\bibitem{BrandtEtAl}
F. Brandt, N. Dragon, M. Kreuzer,
Phys. Lett. B231 (1989) 263;
Nucl. Phys. B332 (1990) 224 and 250

\bibitem{DuboisVioletteEtAl}
M. Dubois--Violette, M. Henneaux, M. Talon, C.--M. Viallet,
Phys. Lett. B267 (1991) 81;
{\it ibid.}\/ B289 (1992) 361

\endbibliography

\bye